\documentclass[12pt]{article}
\usepackage{verbatim,color,amssymb,epsfig}
\usepackage[usenames,dvipsnames]{xcolor}
\usepackage{fancyhdr}
\usepackage{amsmath, amsthm, amssymb}
\usepackage{mathtools}
\usepackage{algorithm,algorithmic}
\usepackage{mathrsfs}
\usepackage{bm}         
\usepackage{enumerate}
\usepackage{multirow}
\usepackage[font={small,it}]{caption}
\usepackage[normalem]{ulem}
\usepackage{url}

\usepackage[authoryear, sort]{natbib}
\bibpunct{(}{)}{;}{a}{,}{,}
\newcommand{\biblist}{
\bibliographystyle{plainnat}
\bibliography{sinha2}
}

\theoremstyle{definition}
\newtheorem{Exa}{Example}
\newcommand\Mhat{{\widehat{MSE}}(\bm{\widehat\mu},\bm{\mu})}
\newcommand\MhatD{{\widehat{MSE}}(\bm{\widehat D},\bm{D})}
\newcommand\NGam{$N\Gamma^{-1}$}
\newcommand\independent{\protect\mathpalette{\protect\independenT}{\perp}}
\def\independenT#1#2{\mathrel{\rlap{$#1#2$}\mkern2mu{#1#2}}}
\DeclareMathOperator*{\argmin}{arg\,min} 

\newcommand*\xbar[1]{%
\hbox{%
\vbox{%
\hrule height 0.5pt 
\kern0.5ex
\hbox{%
\kern-0.2em
\ensuremath{#1}%
\kern-0.1em
}%
}%
}%
}

\begin{document}
\thispagestyle{empty}
\baselineskip=28pt
\begin{center}
{\LARGE{\bf Estimating the Mean and Variance of a High-dimensional
    Normal Distribution Using a Mixture Prior}}
\end{center}
\baselineskip=12pt

\vskip 2mm
\begin{center}
{\bf Shyamalendu Sinha and Jeffrey D.~Hart}\\
\ \\
Department of Statistics, Texas A\&M University, 3143 TAMU, College
Station, TX 77843-3143, ssinha@stat.tamu.edu, hart@stat.tamu.edu\\

\hskip 5mm \\
\end{center}

\begin{abstract}
This paper provides a framework for estimating the mean and variance
of a high-dimensional normal density. The main setting considered is a
fixed number of vector following a high-dimensional normal
distribution with unknown mean and diagonal covariance matrix. The
diagonal covariance matrix can be known or unknown. If the covariance
matrix is unknown, the sample size can be as small as $2$. The
proposed estimator is based on the idea that the unobserved pairs of
mean and variance for each dimension are drawn from an unknown
bivariate distribution, which we model as a mixture of normal-inverse
gammas. The mixture of normal-inverse gamma distributions provides
advantages over more traditional empirical Bayes methods, which are
based on a normal-normal model. When fitting a mixture model, we are
essentially clustering the unobserved mean and variance pairs for each
dimension into different groups, with each group having a different
normal-inverse gamma distribution. The proposed estimator of each mean
is the posterior mean of shrinkage estimates, each of which shrinks a
sample mean towards a different component of the mixture
distribution. Similarly, the proposed estimator of variance has an
analogous interpretation in terms of sample variances and components
of the mixture distribution. If diagonal covariance matrix is known,
then the sample size can be as small as $1$, and we treat the pairs of
known variance and unknown mean for each dimension as random
observations coming from a flexible mixture of normal-inverse gamma
distributions.
\end{abstract}

\baselineskip=12pt
\par\vfill\noindent
\underline{\bf Some Key Words}:
 multivariate normal mean and variance estimation, heteroscedasticity, shrinkage estimator, bivariate density estimation, Dirichlet process mixture model
\par\medskip\noindent
\underline{\bf Short title}: Multivariate Normal Mean and Variance Estimation

\pagenumbering{arabic}
\newlength{\gnat}
\setlength{\gnat}{18pt}
\baselineskip=\gnat

\section{Introduction} \label{sec:sec1}

An old and simple problem in statistics involves estimating the mean of a normal distribution. A somewhat newer and more complex problem is that of estimating the means of many normal distributions when we observe independent samples from these distributions. We consider a version of the latter problem in which $X_{1j},\ldots,X_{nj}$, $j=1,\ldots,q$, are observations generated from the following model:
\begin{align}
\begin{split}
X_{ij}&=\mu_j+\sigma_j \epsilon_{ij}, \quad i=1,\ldots,n, \quad j=1,\ldots,q,\\ 
\epsilon_{ij}& \overset{i.i.d.}{\sim} N(0,1), \quad i=1,\ldots,n, \quad j=1,\ldots,q.\label{eq:LSREeq1}
\end{split}
\end{align}
The following assumptions are made:
\begin{enumerate}[(i)]
 \item The unknown pairs $(\mu_j, \sigma_j^2)$, $j=1,\ldots,q$, are independent and identically distributed and follow an unknown absolutely continuous distribution, denoted by $f_{\mu,{\sigma}^2}$.
 \item The unobserved errors $\epsilon_{ij}$, $i=1,\ldots,n$, $j=1,\ldots,q$, are independent and identically distributed as $f_{\epsilon}$, which is a standard normal density. 
 \item The parameters $(\mu_j, \sigma_j^2)$, $j=1,\ldots,q$, are independent of $\epsilon_{ij}$, $i=1,\ldots,n$, $j=1,\ldots,q$.
\end{enumerate}
The main goal is to estimate $(\mu_j,\sigma_j^2)$, $i=1,\ldots,q$, from $X_{ij}$, $i=1,\ldots,n$, $j=1,\ldots,q$. A secondary goal, which is necessary to efficiently achieve the main goal, is to estimate $f_{\mu,\sigma^2}$, the joint distribution of $(\mu_j,\sigma_j^2)$.

If $\sigma_1^2,\ldots,\sigma_q^2$ are known, replications of the unobserved variable $\mu_j$ are not needed to estimate $\mu_j$. Without loss of generality, we may assume $n=1$, in which case model (\ref{eq:LSREeq1}) reduces to
\begin{align}
\begin{split}
X_{j}&=\mu_j+\sigma_j \epsilon_{j}, \quad j=1,\ldots,q,\\ 
\epsilon_{j}& \overset{i.i.d.}{\sim} N(0,1), \quad j=1,\ldots,q.\label{eq:LSREeq2}
\end{split}
\end{align}
In this case, we observe the pairs $(X_j,\sigma_j^2)$, $j=1,\ldots,q$, and the main goal is to estimate the unknown parameters $\mu_j$, $j=1,\ldots,q$. 

In multivariate notation, $\bm{X}_{i\cdot}={(X_{i1},\ldots,X_{iq})}^T$, $i=1,\ldots,n$, are $n$ observations from a $q$-variate normal distribution with mean vector $\bm{\mu}={(\mu_1,\ldots,\mu_q)}^T$ and variance matrix $\bm{D}=\textrm{Diag}(\sigma_1^2,\ldots,\sigma_q^2)$. 

In the classical one-dimensional framework, i.e.~$q=1$, the sample mean, $\bar{X}_{\cdot 1}=n^{-1}\sum_{i=1}^n X_{i1}$, and $(n+1)^{-1}\sum_{i=1}^n (X_{i1}-\bar{X}_{\cdot 1})^2$ are optimal mean squared error estimators of the population mean and variance, respectively. However, this result does not extend to high-dimensions, as \cite{stein1956inadmissibility} showed that the sample means are inadmissible when $q \ge 3$. The seminal work of \cite{james1961estimation} showed that shrinkage estimators of the means perform better than sample means in terms of mean squared error when $q \ge 3$ and $\sigma_1^2,\ldots,\sigma^2_q$ are all the same (the homoscedastic case) and known. A nice introduction of this class of estimators can be found in the book of \cite{efron2012large}. \cite{efron1973stein} gave an empirical Bayes interpretation of this shrinkage estimator and developed several competing estimators. They noted that even when all $\sigma_j^2$ are known, the James-Stein estimator cannot be extended under heteroscedasticity by simply using the transformation $\sigma_j^{-1} X_{ij}$. This is because the shrinkage factor remains constant under the transformation, as opposed to what intuition entails, namely that more shrinkage should be applied to the components with larger $\sigma_j^2$. They assumed a hierarchical normal model in which $\mu_j\overset{i.i.d.}{\sim} N(0,A)$, and estimated the variance $A$ from the marginal density of $X_{ij}$. As noted by \cite{efron1973stein}, such a hierarchical model is a ``Bayesian statement of belief that the $\mu_j$ are of comparable magnitude,'' a belief which is not always realistic. 

There is a large literature on estimating the mean vector of a multivariate normal distribution under homoscedasticity, using both frequentist and Bayesian approaches. For example, \cite{baranchik1970family} derived the general form of a minimax estimator for the homoscedastic case. \cite{brown1971admissible} derived a general condition for Bayes estimators to be admissible in terms of mean squared error. Using these conditions, \cite{berger1996choice} showed that some common choices of improper prior on hyperparameters lead to inadmissible estimators, and encouraged the use of a proper prior on hyperparameters. \cite{brown2009nonparametric} proposed a nonparametric empirical Bayes solution for estimating the mean.

In contrast, the literature on the heteroscedastic case is scant. \cite{berger1976admissible} provided a minimax estimator when the covariance matrix $\bm{D}$ is known under arbitrary quadratic loss. However, this estimator exhibits the counter-intuitive behavior mentioned before. Recently, there have been a few articles addressing this issue. \cite{xie2012sure} assumed that $\bm{D}$ is known and estimated the mean vector, $\bm{\mu}$, by minimizing Stein's unbiased risk estimate (SURE). They showed that the empirical Bayes maximum likelihood estimator (EBMLE) of $\bm{\mu}$ and SURE estimates of $\bm{\mu}$ do not provide the same solution as in the homoscedastic case and proved a few results about the consistency of the SURE estimates. By not limiting the prior on the normal density, they explored a semiparametric option which we will discuss in detail later. \cite{jing2016sure} further extended the work of \cite{xie2012sure} in the heteroscedastic case when $\bm{D}$ is unknown by modifying the loss function and assuming a gamma prior on the precision parameters, the inverse of the variance parameters.

Theorem 5.7 of \cite{lehmann2006theory} provides a condition for
which the shrinkage estimator becomes a minimax estimator under
squared error loss. However, the family of estimators they
considered applies constant shrinkage to all coordinates, as
opposed to the intuition that components with larger
$\sigma_j^2$ should be shrunk more.
\cite{tan2015improved}
proposed a minimax estimator when the covariance matrix $\bm{D}$ is
known under arbitrary quadratic loss, where the shrinking direction is
open to specification and the shrinking factor is determined. This
minimax estimator is similar to the estimator arising from the
assumption that $\mu_1,\ldots,\mu_q$ are independent with $\mu_j\sim
N(0,A_j)$, $j=1,\ldots,q$. \cite{zhang2017empirical} developed an
empirical Bayes method to estimate a sparse normal
mean. \cite{weinstein2018group} developed an empirical Bayes estimator
assuming that $\sigma_1^2,\ldots,\sigma_q^2$ are part of the random
observations. They binned the pairs $(X_j,\sigma_j^2)$ on the basis of
$\sigma_j^2$ and applied a spherically symmetric estimator separately
in each group. Even though we also assume that $(\mu_j,\sigma_j^2)$
come from a joint distribution, $f_{\mu,\sigma^2}$, our method is
based on modeling the bivariate density of $(\mu,\sigma^2)$ with a
flexible mixture of normal-inverse gamma densities and then estimating
$\bm{\mu}$ and $\bm{D}$.  

\section{Motivation for a New Estimator} \label{sec:sec2}

\subsection{Homoscedastic Case} \label{sec:sec2.1}

Consider the model (\ref{eq:LSREeq1}), where $\sigma_j^2=\sigma^2$, for $j=1,\ldots,q$, and $\sigma^2$ is known, an example discussed in \cite{neyman1948consistent}. We will discuss some existing approaches to estimating $\bm{\mu}$ in this setting and also how our methodology is related to these approaches. Let $\xbar{\bm{X}}$ be the $q$-vector whose $j^{th}$ component is the sample mean $\bar{X}_{\cdot j}=n^{-1} \sum_{i=1}^{n} X_{ij}$, $j=1,\ldots,q$. Then $\xbar{\bm{X}}$ is distributed as $N_q\left(\bm{\mu},\sigma^2\bm{I}/n\right)$. 

\cite{james1961estimation} considered a class of estimators indexed by $c$, which are written as 
\begin{align*}
\bm{\delta}_c^{JS}\left(\xbar{\bm{X}}\right)=\left(1-\frac{\sigma^2}{n} \frac{c}{{\left\lVert \xbar{\bm{X}} \right\rVert}^2}\right)\xbar{\bm{X}},
\end{align*}
where ${\left\lVert \xbar{\bm{X}}\right\rVert}^2={\xbar{\bm{X}}}^{T}\xbar{\bm{X}}$ and the $j^{th}$ element of $\bm{\delta}$, $\delta_j$, is an estimator of $\mu_j$. The average loss, defined by $L(\bm{\delta},\bm{\mu})=q^{-1}{\left\lVert \bm{\delta} - \bm{\mu} \right\rVert}^2$, is used to compare different estimates. \cite{james1961estimation} showed that the constant $c=q-2$ minimizes the risk, $R(\bm{\delta},\bm{\mu})= E_{\bm{\mu}} L(\bm{\delta},\bm{\mu})$, for every $\bm{\mu}$ if $q \geq 3$. We shall call the estimator $\bm{\delta}_{q-2}^{JS}\left(\xbar{\bm{X}}\right)$ simply $\bm{\delta}^{JS}$. \cite{james1961estimation} showed that if $q \geq 3$, $\bm{\delta}^{JS}$ dominates the MLE, $\xbar{\bm{X}}$, in terms of $R(\bm{\delta},\bm{\mu})$ for every choice of $\bm{\mu}$, i.e. $\xbar{\bm{X}}$ is inadmissible. 

\cite{baranchik1970family} considered the following more general family of estimators:
\begin{align*}
\bm{\delta}_r^{JS}\left(\xbar{\bm{X}}\right)=\left(1-\frac{\sigma^2}{n} \frac{r\left({\left\lVert \xbar{\bm{X}} \right\rVert}^2\right)}{{\left\lVert \xbar{\bm{X}} \right\rVert}^2}\right)\xbar{\bm{X}},
\end{align*}
and showed that the estimator is minimax if $r(\cdot)$ is monotone, non-decreasing, and such that $0 \leq r(\cdot) \leq 2(q-2)$. Chapter 5 of \cite{lehmann2006theory} discusses risk properties of these estimators in detail. Another minimax estimator is a version of the James-Stein estimator with non-negative multiplier: 
\begin{align*}
\bm{\delta}^{JS+}\left(\xbar{\bm{X}}\right)=\textrm{max}\left(0, 1-\frac{\sigma^2}{n} \frac{q-2}{{\left\lVert \xbar{\bm{X}} \right\rVert}^2}\right)\xbar{\bm{X}}.
\end{align*}
This estimator dominates the usual James-Stein estimator in terms of $R(\bm{\delta},\bm{\mu})$. All of these shrinkage estimators shrink each coordinate towards $0$. 

\cite{efron1973stein} showed an empirical Bayes connection with the James-Stein estimator by assuming a prior of the form $\mu_j \overset{i.i.d.}{\sim} N(m,\lambda)$, $j=1,\ldots,q$, where $m$ and $\lambda$ are unknown hyperparameters. From Bayes rule, we have that conditional on $\bar{X}_{\cdot 1},\ldots,\bar{X}_{\cdot q},m,\lambda$, 
\begin{align}
\mu_j|\bar{X}_{\cdot j},m,\lambda \overset{indept}{\sim} N\left(\frac{\lambda}{\lambda+\frac{\sigma^2}{n}} \bar{X}_{\cdot j} + \frac{\frac{\sigma^2}{n}}{\lambda+\frac{\sigma^2}{n}} m, \frac{1}{\lambda^{-1}+\frac{n}{\sigma^2}}\right), \quad j=1,\ldots,q \label{eq:homopost}, 
\end{align}
which leads to the shrinkage estimator 
\begin{align*}
\bm{\hat{\mu}} = \xbar{\bm{X}} - \frac{\frac{\sigma^2}{n}}{\lambda+\frac{\sigma^2}{n}}\left(\xbar{\bm{X}}-m\right).
\end{align*}
This estimator is a function of the unknowns $m$ and $\lambda$. These parameters may be estimated using the marginal density
\begin{align*}
\bar{X}_{\cdot j}|m,\lambda \overset{i.i.d.}{\sim} N\left(m,\lambda+\frac{\sigma^2}{n}\right), \quad j=1,\ldots,q,
\end{align*}
from which one may obtain the maximum likelihood estimator (MLE) or method of moments estimator (MOM) of ($m,\lambda)$.

\subsection{Heteroscedastic Case} \label{sec:sec2.2}

When $\sigma_1^2,\ldots,\sigma^2_q$ are not all the same but known, we can modify the James-Stein estimator by using the transformation $\sigma_j^{-1} X_{ij}$, which produces homoscedastic data. Then the James-Stein estimate of $\bm{\mu}$ is 
\begin{align*}
\bm{\delta}^{JS}\left(\xbar{\bm{X}}\right) = \left(1 - \frac{q-2}{n\sum_{j=1}^{q}{\left(\sigma_j^{-1} \bar{X}_{\cdot j}\right)}^2}\right)\xbar{\bm{X}}. 
\end{align*}
As discussed in \cite{efron1973stein} this estimate is not intuitive as we should shrink more those coordinates with larger $\sigma_j^2$. 

When $\sigma_1^2,\ldots,\sigma_q^2$ are not all the same but known, then by assuming the same normal prior that leads to (\ref{eq:homopost}) we obtain 
\begin{align}
\mu_j|\bar{X}_{\cdot j},m,\lambda &\overset{indept}{\sim} N\left(\frac{\lambda}{\lambda+\frac{\sigma_j^2}{n}} \bar{X}_{\cdot j} + \frac{\frac{\sigma_j^2}{n}}{\lambda+\frac{\sigma_j^2}{n}} m, \frac{1}{\lambda^{-1}+\frac{n}{\sigma_j^2}}\right), \quad j=1,\ldots,q \label{eq:heteropost},
\end{align}
which leads to the shrinkage estimator $\xbar{\bm{X}} - \bm{W}\left(\xbar{\bm{X}}-m\right)$, where $\bm{W}$ is a diagonal matrix with $j^{th}$ diagonal element equal to $\frac{\sigma_j^2}{n}{\left(\lambda+\frac{\sigma_j^2}{n}\right)}^{-1}$. To estimate the unknown hyperparameters $m$ and $\lambda$, we may use the marginal density, 
\begin{align*}
X_{ij}|m,\lambda \overset{indept}{\sim} N\left(m,\lambda+\sigma_j^2\right), \quad i=1,\ldots,n, \quad j=1,\ldots,q. 
\end{align*}
However, unlike the homoscedastic case, we cannot estimate $\lambda$ consistently from this marginal density (with $n$ fixed), which impairs the traditional empirical Bayes approach. 

\cite{xie2012sure} addressed this problem which finds a solution of $m$ and $\lambda$ by minimizing SURE, an unbiased estimator of the risk $R(\bm{\delta},\bm{\mu})$. They showed that the SURE estimates are optimal in an asymptotic sense compared to EBMLE or EBMOM. To generalize the estimate, they developed a novel semiparametric approach by not assuming a normal-normal hierarchical model. The semiparametric SURE shrinkage estimator which was discussed in \cite{xie2012sure} assumes that
\begin{eqnarray}\label{eq:SURE}
{\hat{\mu}_j}^{SM} = (1-b_j) \bar{X}_{\cdot j} + b_j m, \quad j=1,\ldots,q. 
\end{eqnarray}
The unbiased estimator of the risk is
\begin{align*}
{SURE}^{SM}(\bm{b},m) = q^{-1} \sum_{j=1}^q \left(b_j^2{\left(\bar{X}_{\cdot j}-m\right)}^2+\left(1-2b_j\right) \frac{\sigma_j^2}{n}\right),
\end{align*}
where $\bm{b}=(b_1,\ldots,b_q)$. The estimator of $\bm{b}$ and $m$ is
\begin{align*}
(\hat{\bm{b}},\hat{m}) = \argmin_{\bm{b},m}{SURE}^{SM}(\bm{b},m),
\end{align*}
subject to $$ 0\le b_j\le1,\quad j=1,\ldots,q, \textrm{ and } b_j \leq
b_l \textrm{ for any $j$ and $l$ such that } \frac{\sigma_j^2}{n}
\leq \frac{\sigma_l^2}{n}.$$ In principle, all of $b_1,\ldots,b_q$
can be distinct if $\sigma_j^2/n \le (\bar{X}_{\cdot j}-m)^2$,
$j=1,\ldots,q$, and $\frac{\sigma_j^2}{n{\left(\bar{X}_{\cdot
      j}-m\right)}^2} \leq
\frac{\sigma_l^2}{n{\left(\bar{X}_{\cdot l}-m\right)}^2}$ in all cases
where $\frac{\sigma_{j}^2}{n} \leq \frac{\sigma_{l}^2}{n}$. If these
conditions do not hold, the number of distinct $b_j$ reduces. In
practice, the number of distinct $b_j$ is very low compared to $q$
since $\textrm{Prob}\left(\sigma_j^2(\bar{X}_{\cdot l}-m)^2 >
\sigma_l^2(\bar{X}_{\cdot j}-m)^2\right)$ is often relatively large
even if $\sigma_j^2 < \sigma_l^2$. A natural extension of SURE
minimization, where all of $m_1,\ldots,m_q$ are distinct, is not
possible because the solution will be $m_j = \bar{X}_{\cdot j}$,
i.e. $b_j=0$, leading to a non-shrinkage estimator. The approach of
\cite{xie2012sure} is tantamount to assuming that $\mu_1,\ldots,\mu_q$
are drawn from a mixture of normals that are all centered at $m$ but
have different variances. 

The approach of \cite{xie2012sure} is less general than the one
considered in the current paper where we consider a mixture
distribution whose components can have different means {\it and}
variances. Approaches that use the same shrinkage for each
  component and/or the assumption that the components of the mean
  vector follow a unimodal distribution can produce very poor
  estimates. This will occur, for example, when the $\mu_j$s come from
  a bimodal distribution with widely separated modes. Our approach
  based on a \NGam\ mixture mitigates this problem by using ``local''
  shrinkage, i.e., shrinkage of a sample mean towards the mixture
  component to which its $\mu_j$ is most likely to belong.

\cite{weinstein2018group} proposed a group-linear empirical Bayes
method, which treats known variances as part of the random
observations and applies a spherically symmetric estimator to each
group separately. This shrinks sample means in different directions,
but their clustering mechanism only depends on $\sigma_j^2$. This is
unrealistic as the shrinkage directions should depend on the modes of
the distributions of the unobserved $\mu_j$, and the shrinkage factors
should depend on the known $\sigma_j^2$. If $\mu_j$ is a smooth
function of $\sigma_j^2$, group-linear algorithms perform well as the
clustering by similar $\log(\sigma_j^2)$ means unobserved values of
$\mu_j$ in the same cluster are also similar. However, if $\mu_j$ and
$\sigma_j^2$ are independent, clustering by group-linear algorithms is
not effective, resulting in poor estimates compared to SURE estimates.  

\cite{weinstein2018group} obtained results for the heteroscedastic case where $\sigma_1^2,\ldots,\sigma^2_q$ are i.i.d. Likewise, our proposed estimate assumes that $\sigma_1^2,\ldots,\sigma_q^2$ are i.i.d., but it has at least two practical advantages over that of \cite{weinstein2018group}. First of all, we need not assume that $\sigma_1^2,\ldots,\sigma_q^2$ are known, and secondly no binning of $\sigma_1^2,\ldots,\sigma_q^2$ (with the attendant problem of choosing the number of bins) is required. We model the joint density of $(\mu_j, \sigma_j^2)$ by a flexible mixture of normal-inverse gamma distributions. As we will show later, our estimators of $\mu_j$ are similar in form to the SURE estimate (\ref{eq:SURE}), but, when appropriate, they shrink $\bar{X}_{\cdot j}$ towards the mean of a mixture component rather than towards the overall mean. This has the potential of producing better estimates of $\mu_1,\ldots,\mu_q$ when the distribution of $\mu_j$ is nonnormal.

\cite{jing2016sure} extended the result from \cite{xie2012sure} to the case where $\sigma_1^2,\ldots,\sigma_q^2$ are unknown. They used a different risk function, $$q^{-1}\sum_{j=1}^q E_{\bm{\mu},\bm{D}} \left({(\hat{\mu}_j-\mu_j)}^2 + n^{-2}{(\hat{\sigma}_j^2-\sigma_j^2)}^2\right),$$ and then minimized unbiased estimators of it by shrinking sample mean and sample variance, $\bar{X}_{\cdot j} $ and $S_{\cdot j}^2$ respectively, where $S_{\cdot j}^2={(n-1)}^{-1}\sum_{i=1}^n (X_{ij}$ $-\bar{X}_{\cdot j})^2$, towards appropriate direction. However, they used constant shrinkage factors for estimating each of $\mu_j$ and $\sigma_j^2$. Our method naturally extends to the case where $\sigma_1^2,\ldots,\sigma_q^2$ are unknown. This is more general than assuming a normal-normal hierarchical model as the mixture of normals provides a more flexible prior compared to using a single normal. Each component has a different mean and we shrink each $\mu_j$ in an appropriate direction rather than one general direction, which was a main drawback in all previous works. 

\section{Modeling the Joint Distribution of $\mu$ and $\sigma^2$ by a Mixture of Normal-Inverse Gamma Distributions} \label{sec:sec3} 

To estimate the bivariate density $f_{\mu,\sigma^2}$ nonparametrically, it is reasonable to use a mixture of bivariate densities, which underlies most mainstream approaches of density estimation, including kernel techniques (\cite{silverman1986density}), nonparametric maximum likelihood (\cite{lindsay1983geometry}), and Bayesian approaches using mixtures induced by a Dirichlet process (\cite{ferguson1983bayesian} and \cite{escobar1995bayesian}). 

In this paper, we define gamma and inverse-gamma densities as
\begin{align*}
 G(x|a,b) = \frac{b^a}{\Gamma(a)} x^{a-1} e^{-b x}I_{(0,\infty)}(x), \quad IG(x|a,b) = \frac{b^a}{\Gamma(a)} x^{-a-1} e^{-b/x}I_{(0,\infty)}(x),
\end{align*}
respectively, where $\Gamma$ is the gamma function and $I_A$ is the indicator function for the set $A$. Though it is more common to use a mixture of normal densities, $\sigma^2$ has support only on the positive side of the real line, and hence using a mixture of bivariate normals seems unreasonable. An easy way to get around the problem of positive support is to estimate the density of $\log(\sigma^2)$ using a mixture of normals. However, if we assume $f_{\epsilon}$ is standard normal, then a mixture of bivariate normals for the joint density of $(\mu,\log(\sigma^2))$ is not a conjugate prior. A mixture of normal-inverse-gamma (N$\Gamma^{-1}$) densities seems more reasonable as the posterior density of the parameters of interest belongs to a known family of densities. A N$\Gamma^{-1}(m,\lambda,\alpha,\beta)$ density has two components, normal and inverse-gamma, and is defined by
\begin{align*}
g(\mu,\sigma^2|m,\lambda,\alpha,\beta) = N(\mu|m,\sigma^2/\lambda) IG(\sigma^2|\alpha,\beta),
\end{align*}
where $N(\cdot|m,s^2)$ denotes a normal density function with mean $m$ and variance $s^2$. 

The density $f_{\mu,\sigma^2}$ is defined to be a mixture of \NGam\ densities induced by a Dirichlet process with concentration parameter $\gamma$. Let $\bm{\pi}$ denote the vector of random mixture weights. \cite{sethuraman1994constructive} describes the {\it stick-breaking} process, a method to construct $\bm{\pi}=\{\pi_k\}_{k=1}^{\infty}$ so that $\sum_{k=1}^\infty \pi_k=1$. For $r=1,2,\ldots$, the process can be described as
\begin{align*}
\pi_r = s_r \prod_{j=1}^{r-1}(1-s_j), \quad s_1,s_2,\ldots \overset{i.i.d.}{\sim} \textrm{Beta}(1,\gamma),
\end{align*}
where $\textrm{Beta}(a,b)$ denote a Beta distribution with parameters $(a,b)$. Let us denote this process as $\textrm{Stick}(\,\cdot\,|\gamma)$. The quantities $\bm{m}$, $\bm{\lambda}$, $\bm{\alpha}$ and $\bm{\beta}$ are the vectors of parameters of the N$\Gamma^{-1}$ densities that make up the mixture. Let $\bm{\Theta}=[\bm{m},\bm{\lambda},\bm{\alpha},\bm{\beta}]$ be a matrix of four columns whose $r^{th}$ row, $\bm{\Theta_r}$, contains parameters for the $r^{th}$ component of the mixture. The Dirichlet process mixture model (DPMM), denoted $DP(\gamma,G_0)$ with concentration parameter $\gamma$, base measure $G_0$ and \NGam\ mixture components, is specified as 
\begin{align*}
f_{\mu,\sigma^2}(\mu,\sigma^2|\bm{\Theta},\bm{\pi}) = \sum_{r=1}^{\infty} \pi_r g(\mu,\sigma^2|\bm{\Theta_r}),\quad \bm{\Theta_r} \overset{i.i.d.}{\sim} G_0(\,\cdot\,|\bm{\Theta_{H}}), \quad \bm{\pi} \sim \textrm{Stick}(\,\cdot\,|\gamma).
\end{align*}
The prior $G_0$ for the component parameters is taken to be as follows: $m_r$, $\lambda_r$, $\alpha_r$, and $\beta_r$ are independent with
\begin{align*}
m_r\sim N(m_0,\zeta^2),\quad \lambda_r \sim G(a_{\lambda},b_{\lambda}), \quad \alpha_r \sim G(a_{\alpha},b_{\alpha}), \quad \beta_r \sim G(a_{\beta},b_{\beta}).
\end{align*}
The distribution $G_0(\,\cdot\,|\bm{\Theta_H}$ depends on $\bm{\Theta_H}$, the vector of all hyperparameters $(m_0,\zeta^2,a_{\lambda},b_{\lambda},a_{\alpha},b_{\alpha},a_{\beta},b_{\beta})$.

Even though the mixture model theoretically has a countably infinite
number of components, given a dataset, one can only use a mixture
model with a finite number of components. Indeed, in practice, a
finite number of components is adequate. \cite{ishwaran2001gibbs}
constructed a useful class of truncated Dirichlet processes, denoted
$DP_k(\gamma, G_0)$, by applying truncation to standard Dirichlet
processes, where the number of components is fixed at $k$. The
truncation is applied by assuming $\pi_{k+1}=\pi_{k+2}=\ldots=0$ and
replacing $\pi_k$ by $1-\sum_{r=1}^{k-1} \pi_r$. They showed that the
expected sum of moments of discarded random weights decreases
exponentially fast in $k$, and thus, for a moderate $k$, we should be
able to achieve an accurate
approximation. \cite{rousseau2011asymptotic} discussed behavior
  of overfitted mixtures and showed that carefully chosen priors tend
  to empty the extra components, thus mitigating the overfitting
  effect of the DP. We shall use $DP_k(\gamma,G_0)$ in order to model
the density $f_{\mu,\sigma^2}$.  

Since we have measurement error, we do not observe the pair
$(\mu_j,\sigma_j^2)$ directly. Instead, we observe
${\{X_{ij}\}}_{i=1}^{n}$, which will be referred to as $\bm{X}_{\cdot
  j}$, a vector of observed replications of the true unobserved
variable $\mu_j$. As we already assumed the error density to be
standard normal, the joint density of $\bm{X}_{\cdot j}$ given $\mu_j$
and $\sigma_j^2$ is 
\begin{align*}
f(\bm{X}_{\cdot j}|\mu_j,\sigma_j^2)=\prod_{i=1}^n\frac{1}{\sigma_j}f_{\epsilon}\left(\frac{X_{ij}-\mu_j}{\sigma_j}\right)=\prod_{i=1}^n \frac{1}{\sqrt{2\pi}\sigma_j} e^{-\frac{1}{2\sigma_j^2}{(X_{ij}-\mu_j)}^2} .
\end{align*}
Let $Z_j$ be a latent variable indicating the component of the mixture distribution from which the pair $(\mu_j,\sigma_j^2)$ was drawn. The conditional joint density of $(\mu_j,\sigma_j^2)$ is
\begin{align*}
f(\mu_j,\sigma_j^2|\bm{\Theta},Z_j=z_j)=g(\mu_j,\sigma_j^2|\bm{\Theta_{z_j}}).
\end{align*}
The prior probability mass function (p.m.f.) of the latent variable $Z_j$ is 
\begin{align*}
p(Z_j=z_j|\bm{\pi})=\pi_{z_j}. 
\end{align*}

Let ${\bf X}$ denotes the all $n \times q$ observations, $\bm{X}_{\cdot 1},\ldots,\bm{X}_{\cdot q}$.
Let $\mathbb{U}_r=\{j:Z_j=r\}$, and $c_r$ be the cardinality of $\mathbb{U}_r$.
Also, We make the following assumptions:
\begin{enumerate}[(i)]
\item ${\bf X} \independent Z_1,\ldots,Z_q,\bm{\Theta},\bm{\pi} | \bm{\mu}, \bm{D}$,
\item The conditional distribution of $\bm{\mu}, \bm{D}$ given $\bm{\Theta},\bm{\pi},Z_1=z_1,\ldots,Z_q=z_q$ is $$
\prod_{j=1}^q g(\mu_j,\sigma_j^2|\bm{\Theta}_{z_j})$$,
\item $Z_1,\ldots,Z_q \independent \bm{\Theta} | \bm{\pi}$,
\item $\bm{\Theta} \independent \bm{\pi}$,
\end{enumerate}
where $U \independent V|W$ denotes that two random variables $U$ and $V$ are independent conditional on $W$. The posterior is proportional to 
\begin{align*}
& f({\bf X}|\bm{\mu}, \bm{D}, Z_1, \ldots, Z_q, \bm{\Theta}, \bm{\pi}) f(\bm{\mu}, \bm{D}, Z_1, \ldots, Z_q, \bm{\Theta}, \bm{\pi})\\
& \quad = \prod_{j=1}^q f(\bm{X}_{\cdot j}|\mu_j,\sigma_j^2)
  \prod_{j=1}^q g(\mu_j,\sigma_j^2|\bm{\Theta}_{z_j}) \prod_{r=1}^k
       {\pi_{r}}^{c_r} \prod_{r=1}^k
       G_0(\bm{\Theta_r}|\bm{\Theta_{H}}) \textrm{Dir}(\bm{\pi}|\gamma
       \bm{1}_k/k), 
\end{align*}
where Dir and $\bm{1}_k$ denote the Dirichlet distribution and a
  $k$-dimensional vector of $1$s, respectively.  We may reparametrize
$\alpha_r$ and $\beta_r$ in terms of location and scale parameters. If
$\delta_r$ denotes a point between the mean and mode of the
$IG(\alpha_r,\beta_r)$, then we can rewrite the rate parameter
$\beta_r$ as $\delta_r\alpha_r$. The quantities $\delta_r$ and
$\alpha_r$ can be treated as location and scale parameters
respectively. Since $\delta_r$ is the location parameter of a density
with positive support, we can use a gamma prior on $\delta_r$ just as
we did for $\beta_r$ with shape parameter $a_{\delta}$ and scale
parameter $b_{\delta}$.  

\subsection{Algorithm to Estimate Unknown Parameters}\label{sec:3.1}

We will find estimates of the parameters $(\bm{\Theta},\bm{\pi})$ by using an MCMC algorithm to approximate their posterior density. In the notation that follows, $\theta|\cdot$ stands for the conditional distribution of $\theta$ given the data and all unknowns besides $\theta$. The full conditional posterior densities of $\mu_j$ and $\sigma_j^2$ are normal and inverse-gamma, respectively. The full conditional posterior densities of $m_r$, $\lambda_r$, and $\bm{\pi}$ follow normal, gamma, and Dirichlet densities, respectively. The parameters, $\alpha_r$ and $\beta_r$ do not have a standard density. Therefore, we use a Metropolis-Hastings algorithm to sample from these densities. 

\begin{algorithm*}
\caption{MCMC Algorithm to Estimate $(\mu, \sigma^2)$}
\begin{algorithmic}[1]
\small
\STATE Standardize the data $X_{ij} = \left(X_{ij}-\bar{\bar{X}}\right)/S$, where $\bar{\bar{X}}$ and $S$ is the grand mean and grand standard deviation, respectively.
\STATE Run $k$-means clustering on $(\bar{X}_{\cdot j}, S^2_{\cdot j}), \quad j=1,\ldots,q$.
\STATE Initialize $Z_j$, $j=1,\ldots,q$ with the values that indicate the cluster in which $(\bar{X}_{\cdot j}, S^2_{\cdot j})$ belongs.
\STATE Initialize $m_r$, $r=1,\ldots,k$ with the centers of $\bar{X}_{\cdot j}$ clusters from the $k$-means output.
\STATE Initialize $\lambda_r, \alpha_r, \beta_r$, $r=1,\ldots,k$ all with 1.
\STATE Initialize $l=1$ and start the MCMC chain with with these initial values.
\STATE Generate $\mu_j$ form $N\left(\frac{n\bar{X}_{\cdot j}+m_{z_j}\lambda_{z_j}}{n+\lambda_{z_j}},\frac{\sigma_j^2}{n+\lambda_{z_j}}\right)$, for $j=1,\ldots,q$.
\STATE Generate $\sigma_j^2$ from $IG\left(\frac{n+1}{2}+\alpha_{z_j},\frac{1}{2}\sum_{i=1}^{n}{\left(X_{ij}-\mu_j\right)}^2+\frac{\lambda_{z_j}}{2}{\left(\mu_j-m_{z_j}\right)}^2 + \beta_{z_j}\right)$, for $j=1,\ldots,q$.
\STATE Generate $Z_j$ such that $P(Z_j=r) \propto \pi_{r} g\left(\mu_j,\sigma_j^2|m_{r},\lambda_{r},\alpha_{r},\beta_{r}\right)$, for $r=1,\ldots,k$, $j=1,\ldots,q$.
\STATE Generate $m_r$ from $N\left(\frac{m_0\zeta^{-2} + \lambda_r \sum_{j \in \mathbb{U}_r} \mu_j\sigma_j^{-2}}{\zeta^{-2} + \lambda_r \sum_{j \in \mathbb{U}_r} \sigma_j^{-2}}, \frac{1}{\zeta^{-2} \lambda_r \sum_{j \in \mathbb{U}_r} \sigma_j^{-2}}\right)$, for $r=1,\ldots,k$.
\STATE Generate $\lambda_r$ from $G\left(\frac{c_r}{2}+a_{\lambda}, \sum_{j \in \mathbb{U}_r} \frac{{(\mu_j-m_r)}^2}{2 \sigma_j^2} + b_{\lambda}\right)$, for $r=1,\ldots,k$.
\STATE Generate $\alpha_r$ from $\frac{{b_{\alpha}}^{a_{\alpha}}}{\Gamma(a_{\alpha})} {\alpha_{r}}^{a_{\alpha}-1} e^{-b_{\lambda}\alpha_{r}} \prod_{j \in \mathbb{U}_r} \frac{{\beta_{r}}^{\alpha_{r}}}{\Gamma(\alpha_r)} {(\sigma_j^2)}^{-\alpha_{r}-1} e^{-\beta_{r}\sigma_j^{-2}}$, for $r=1,\ldots,k$ using Metropolis~-~Hastings algorithm.
\STATE Generate $\beta_r$ from $\frac{{b_{\beta}}^{a_{\beta}}}{\Gamma(a_{\beta})} {\beta_{r}}^{a_{\beta}-1} e^{-b_{\beta}\beta_{r}} \prod_{j \in \mathbb{U}_r} \frac{{(\beta_{r})}^{\alpha_{r}}}{\Gamma(\alpha_r)} {(\sigma_j^2)}^{-\alpha_{r}-1} e^{-\beta_{r}\sigma_j^{-2}}$, for $r=1,\ldots,k$ using Metropolis~-~Hastings algorithm.
\STATE Generate $\bm{\pi}$ from $\textrm{Dir}\left(c_1+\frac{\gamma}{k},\ldots,c_k+\frac{\gamma}{k}\right)$.
\STATE $l=l+1$ and if $l < s$ go to step 6.
\end{algorithmic}
\end{algorithm*}

The $l^{th}$ full iteration of the MCMC algorithm produces values $\mu_{1l},\ldots,\mu_{ql}$, $\sigma_{1l}^2,\ldots,\sigma_{ql}^2$. Our estimates of $\mu_1,\ldots,\mu_q$, $\sigma_1^2,\ldots,\sigma_q^2$ are obtained by averaging $\mu_{1l},\ldots,\mu_{ql}$, $\sigma_{1l}^2,\ldots,\sigma_{ql}^2$ over all iterations, which of course provides an approximation to the posterior mean of each parameter. Furthermore, at every MCMC iteration we obtain $\bm{\Theta}_l$ and $\bm{\pi}_l$, from which we can calculate values of the mixture density over a grid. Averaging density values over all iterations leads to an estimate of $f_{\mu,\sigma^2}$.

Denote our estimate of $\bm{\mu}$ by $\bm{\hat\mu}^{DPMM}$, where $DPMM$ stands for Dirichlet process mixture model. The $j^{th}$ component of $\bm{\hat\mu}^{DPMM}$, $\hat\mu_j^{DPMM}$, approximates $E(\mu_j|\textrm{data})$. Defining $$\hat\mu(z_j,m_{z_j},\lambda_{z_j})=(n\bar{X}_{\cdot j}+m_{z_j}\lambda_{z_j})/(n+\lambda_{z_j}),$$ the conditional posterior density of $\mu_j$, and iterated expectation imply that $$ E(\mu_j|\textrm{data})=E\left[\hat\mu(z_j,m_{z_j},\lambda_{z_j})|\textrm{data}\right]$$. Letting $b_j=\lambda_{z_j}/(n+\lambda_{z_j})$, we have $$\hat\mu(z_j,m_{z_j},\lambda_{z_j})=(1-b_j)\bar{X}_{\cdot j}+b_j m_{z_j},$$ and so for each choice of the unknown parameters $(z_j,m_{z_j},\lambda_{z_j})$, $\hat\mu(z_j,m_{z_j},\lambda_{z_j})$ is a shrinkage estimate having the same form as the SURE estimates (\ref{eq:SURE}). The actual estimate of $\mu_j$, $E(\mu_j|\textrm{data})$, is simply the posterior mean of all these shrinkage estimates. In the event that $\mu_j$ comes from, say, component 1 with high probability $${\hat{\mu}_j}^{DPMM} \approx nE((n+\lambda_1)^{-1}|\textrm{data},Z_j=1) \bar{X}_{\cdot j} + E(m_1\lambda_1(n+\lambda_1)^{-1}|\textrm{data},Z_j=1),$$ and hence $\bar{X}_{\cdot j}$ shrinks towards the posterior mean of $m_1$ rather than the overall mean. Certainly, in cases where the distribution of $\mu_j$ is multimodal with widely separated modes this scheme should produce much better estimates of $\bm{\mu}$ than does (\ref{eq:SURE}), a claim confirmed by simulations in Sections \ref{sec:sec4.1}-\ref{sec:sec4.2}. From the full conditional distribution of $\sigma_j^2$s the posterior mean of $\sigma_j^2$ is 
\begin{eqnarray}\label{eq:sighat2}
E(\sigma_j^2|\textrm{data})=E\left\{\frac{(n-1)\tilde\sigma_j^2+2\alpha_{z_j}(\beta_{z_j}/\alpha_{z_j})}{n-1+2\alpha_{z_j}}\Bigg|\textrm{data}\right\},
\end{eqnarray}
where $$\tilde\sigma_j^2=(n-1)^{-1}\left[(n-1)S_{\cdot j}^2+n(\bar{X}_{\cdot j}-\mu_j)^2+\lambda_{z_j}(\mu_j-m_{z_j})^2\right].$$ So, $E(\sigma_j^2|\textrm{data})$ has an interpretation analogous to that of $E(\mu_j|\textrm{data})$. The quantity $\beta_{z_j}/\alpha_{z_j}$ may be regarded as a location parameter of the inverse-gamma component as it lies between the mode and the mean, and therefore $E(\sigma_j^2|\textrm{data})$ is the posterior mean of shrinkage estimates each of which shrinks the variance estimate $\tilde\sigma_j^2$ towards $\beta_{z_j}/\alpha_{z_j}$.

\subsection{Choice of Prior Parameters} \label{sec:sec3.2}

We can run a fully Bayes approach using a prespecified value of $\bm{\Theta_{H}}$ and a non-informative prior on $\bm{\Theta}$, or take an empirical Bayes approach to estimate $\bm{\Theta_H}$ from the data. 
Even though we do not observe $(\mu_j,\sigma_j^2)$ directly, we can perceive the problem as one of clustering the $(\mu_j,\sigma_j^2)$ pairs, where each cluster has a different N$\Gamma^{-1}$ density. The parameter $m_r$ denotes the mean of all $\mu_j$ that belong to the $r^{th}$ cluster. The parameters $m_0$ and $\zeta^2$ are the mean and variance of each $m_r$. Let $\bar{\bar{X}}={(nq)}^{-1}\sum_{i=1}^{n} \sum_{j=1}^{q}  X_{ij}$ denote the grand mean and $S^2={(nq-1)}^{-1} \sum_{i=1}^{n}\sum_{j=1}^{q}(X_{ij}-\bar{\bar{X}})^2$ the grand variance. It is reasonable to estimate $m_0$ with its unbiased estimator, the grand mean $\bar{\bar{X}}$. Note that 
\begingroup\makeatletter\def\f@size{10}\check@mathfonts
\begin{align*}
 & E(X_{ij}|\bm{\Theta},\bm{\pi})=E(E(X_{ij}|\mu_j,\sigma_j^2)|\bm{\Theta},\bm{\pi})=E(\mu_j|\bm{\Theta}, \bm{\pi})=\sum_{r=1}^k \pi_r m_r,\\
\quad & E(X_{ij}|\bm{\Theta_H},\gamma)=E(E(X_{ij}|\bm{\Theta},\bm{\pi})|\bm{\Theta_H},\gamma)=E\left(\sum_{r=1}^k \pi_r m_r|\bm{\Theta_H},\gamma\right)=m_0.
\end{align*}\endgroup
On the other hand, estimating $\zeta^2$ is more difficult as the conditional variance of the sample means depends on $\zeta^2$ and many other parameters. Note that 
\begingroup\makeatletter\def\f@size{10}\check@mathfonts
\begin{align}
\begin{split}
\textrm{var}(\bar{X}_{\cdot j}|Z_i=r,\bm{\Theta}) &= \textrm{var}(E(\bar{X}_{\cdot j}|\mu_j,\sigma_j^2)|Z_j=r,\bm{\Theta}) + E(\textrm{var}(\bar{X}_{\cdot j}|\mu_j,\sigma_j^2)|Z_j=r,\bm{\Theta})\\
&= \textrm{var}(\mu_j|Z_j=r,\bm{\Theta}) + n^{-1}E(\sigma_j^2|Z_j=r,\bm{\Theta})\\ \label{eq:hyperpar1}
&= \frac{\beta_r}{\lambda_r(\alpha_r-1)} + \frac{\beta_r}{n(\alpha_r-1)}=\frac{\beta_r}{(\alpha_r-1)}\left(\frac{1}{n}+\frac{1}{\lambda_r}\right),\\
\textrm{var}(\bar{X}_{\cdot j}|\bm{\Theta},\bm{\pi}) &= \textrm{var}(E(\bar{X}_{\cdot j}|Z_j=r,\bm{\Theta}_r)|\bm{\Theta},\bm{\pi}) + E(\textrm{var}(\bar{X}_{\cdot j}|Z_j=r,\bm{\Theta}_r)|\bm{\Theta},\bm{\pi})\\
&= \textrm{var}(m_r|\bm{\Theta},\bm{\pi}) + E\left(\frac{\beta_r}{(\alpha_r-1)}\left(\frac{1}{n}+\frac{1}{\lambda_r}\right)|\bm{\Theta},\bm{\pi}\right)\\
&= \sum_{r=1}^{k} \pi_r m_r^2 - {\left(\sum_{r=1}^{k} \pi_r m_r\right)}^2 + \sum_{r=1}^{k} \frac{\pi_r \beta_r}{(\alpha_r-1)}\left(\frac{1}{n}+\frac{1}{\lambda_r}\right),\\
\textrm{var}(\bar{X}_{\cdot j}|\bm{\Theta_H},\gamma) &= \textrm{var}(E(\bar{X}_{\cdot j}|\bm{\Theta},\bm{\pi})|\bm{\Theta_H},\gamma) + E(\textrm{var}(\bar{X}_{\cdot j}|\bm{\Theta},\bm{\pi})|\bm{\Theta_H},\gamma)\\
 &>  \textrm{var}\left(\sum_{r=1}^{k} m_r \pi_r|\bm{\Theta_H},\gamma\right)\\ &=\frac{2m_0^2\gamma(k-1)}{\gamma+1} + \frac{\zeta^2(k\gamma+1)}{\gamma+1} \geq \zeta^2.
 \end{split}
\end{align}\endgroup
The inequality in the last line of (\ref{eq:hyperpar1}) is intuitively clear as $\zeta^2$ can be seen as the between group variance of $\mu_j$, which must be less than the total variance of $\mu_j$. We will use $S^2_{\bar{X}}={(q-1)}^{-1}\sum_{j=1}^q {(\bar{X}_{\cdot j}-\bar{\bar{X}})}^2$ as our choice of $\zeta^2$ in the prior for $m_r$. Doing so is somewhat informative, but not too informative since $S^2_{\bar{X}}$ estimates $\textrm{var}(\bar{X}_{\cdot j}|\bm{\Theta}_{\bm{H}},\gamma)$, which is larger than $\zeta^2$.

An important parameter of the N$\Gamma^{-1}$ mixtures is $\lambda_r$, whose prior has two hyperparameters, $a_{\lambda}$ and $b_{\lambda}$. From conditional posterior density of $\mu_j$s, we can interpret $\lambda_{z_j}$ as a shrinkage parameter. If $\lambda_{z_j}$ tends to $0$ then the posterior density of $\mu_j$ is centered at the sample mean. The quantity $\lambda_{z_j}$ controls the amount of shrinkage towards the mean of the mixture component. Also, We have
\begin{align*}
\frac{E(\sigma_j^2|Z_j=r,\bm{\Theta}_r)}{\textrm{var}(\mu_j|Z_j=r,\bm{\Theta}_r)} = \lambda_r, 
\end{align*}
which means that $\lambda_r$ may be regarded as a noise to signal ratio. In many, if not most, cases one anticipates that noise to signal ratios will be smaller than 1, which motivates choosing $a_\lambda$ and $b_\lambda$ to produce values of $\lambda_r$ that are smaller than 1 with fairly high probability. 

The prior on mixing probabilities $\bm{\pi}$ is a Dirichlet density
with parameter $\gamma$. \cite{ferguson1983bayesian} discussed in
detail two independent interpretations of the Dirichlet process
parameter $\gamma$. The first one concerns the relative size of
$\pi_r$ and the second one concerns prior information. A smaller value
of $\gamma$ means there are big differences in $\pi_r$ values and also
that we mistrust our prior. So, posterior estimates will be strongly
influenced by the data. \cite{rousseau2011asymptotic} studied the
  behavior of the posterior distribution for overfitted mixture models
  when the data are observed without error.
They proved
  that, under a few mild assumptions, if $\gamma/k < 2$, the posterior
  distribution of $(\mu,\sigma^2)$ has stable behaviour. Our
    situation is somewhat different in that the variables that follow
    a mixture model are observed with error. Nonetheless, we will
    follow the advice of \cite{rousseau2011asymptotic} and use
    $\gamma=0.1$ in most of our simulations and all of our data 
    applications. In simulations 
    reported at the end of Section \ref{sec:sec4.2}, we found that larger
    values of $\gamma$ tend to yield more components than the true
  number. However, we also found that having extra components in the
  mixture model has little effect on the quality of estimates of $\mu_j$ and
  $\sigma_j^2$, at least when $k << q$. 
  
Let $\hat{\sigma}_j^2=(n-1) S_{\cdot j}^2/n$ denote an estimate of $\sigma_j^2$. Now,
\begin{align}
\begin{split}
(nq-1)S^2 &= (n-1) \sum_{j=1}^q S_{\cdot j}^2 + n(q-1) S^2_{\bar{X}},\\
\frac{nq-1}{nq}S^2 &= q^{-1}\sum_{j=1}^q \frac{n-1}{n} S_{\cdot j}^2 + \frac{q-1}{q} S^2_{\bar{X}},\\
S^2 &\approx  q^{-1}\sum_{j=1}^q \hat{\sigma}_j^2 + S^2_{\bar{X}} \label{varsplit}, \quad \textrm{if } q \rightarrow \infty \textrm{ and $n$ is fixed}.
\end{split}
\end{align}

We can rewrite model (\ref{eq:LSREeq1}) as
\begin{align*}
X^{'}_{ij} = \frac{X_{ij}-\bar{\bar{X}}}{S} = \frac{\mu_j-\bar{\bar{X}}}{S} + \frac{\sigma_j}{S} \epsilon_{ij} = \mu^{'}_j + \sigma^{'}_j \epsilon_{ij},
\end{align*}
where $\mu^{'}_j=\left(\mu_j-\bar{\bar{X}}\right)/S$ and
$\sigma_j^{2'}=\sigma_j^2/S^2$. If 
$$
(\mu_j,\sigma_j^2)|Z_j=r \sim
N{\Gamma}^{-1}(m_r,\lambda_r,\alpha_r,\beta_r),
$$
then $$(\mu_j^{'},\sigma_j^{2'})|Z_j=r \sim
N{\Gamma}^{-1}(m_r^{'},\lambda_r,\alpha_r,\beta_r^{'}),$$ where
$m_r^{'}=\left(m_r-\bar{\bar{X}}\right)/S$ and
$\beta_r^{'}=\beta_r/S^2$. Similarly, the new prior is such that
$m_r^{'} \sim N(m_0^{'}, \zeta^{2'})$ and $\beta^{'} \sim
G(a_\beta,b_\beta^{'})$, where $m_0^{'}=\left(m_0 -
\bar{\bar{X}}\right)/S$, $\zeta^{2'}=\zeta^2/S^2$, and
$b_\beta^{'}=b_\beta S^2$.

For the standardized data, as $\textrm{var}(X^{'}_{ij})=1$, equation (\ref{varsplit}) implies that
the average of estimated $\sigma_j^{2'}$ cannot be more than 1. The quantities,
$a_{\alpha}$, $b_{\alpha}$, $a_{\beta}$, and $b_{\beta}^{'}$ are the
hyperparameters for $\alpha_r$ and $\beta_r^{'}$, the scale and rate
parameters of the inverse-gamma distributions comprising the
mixture. We may choose the hyperparameters for the standardized data
in such a way that the prior for $\alpha_r$ and $\beta_r$ has low
information. For all applications in this paper, we used
$a_{\alpha}=b_{\alpha}=a_{\beta}=b_{\beta}^{'}=1$. If one is
interested in using an even less informative prior, they may choose
these parameters to be, say, 0.1
or 0.01 instead of 1. We experimented with these hyperparameter values,
and found that, at least in all the cases considered in this paper,
they had little impact on mean squared error. For model (\ref{eq:LSREeq2}), we used $a_{\alpha}=a_{\beta}=1$, $b_{\alpha}=\textrm{var}\left(\sigma_j^{-2'}\right)/{\left(E\left(\sigma_j^{-2'}\right)\right)}^2$, and $b^{'}_{\beta}=\textrm{var}\left(\sigma_j^{-2'}\right)/E\left(\sigma_j^{-2'}\right)$.

\section{Simulation Study} \label{sec:sec4}

In this section, we conduct a number of simulations to compare different estimates of estimating $\bm{\mu}$ and $\bm{D}$. We simulated data from either model (\ref{eq:LSREeq1}) or model (\ref{eq:LSREeq2}) using a number of different choices for $f_{\mu,\sigma^2}$. To evaluate an estimator $\bm{\widehat\mu}$ of $\bm{\mu}$, we approximate the following version of mean squared error: $$ MSE(\bm{\widehat\mu},\bm{\mu})=E\left[\frac{1}{q}\sum_{j=1}^q(\hat\mu_j-\mu_j)^2\right],$$ where the expectation is taken with respect to the joint distribution of ${\bf X}$ given $\bm{\Theta},\bm{\pi}$. In using this risk function we are taking into account randomness due to $(\mu_j,\sigma_j^2)$, $j=1,\ldots,q$. In our simulation study, each new data set is obtained by generating new values $(\mu_j,\sigma_j^2,\bm{\epsilon}_{\cdot j})$, $j=1,\ldots,q$, where $\bm{\epsilon}_{\cdot j}=(\epsilon_{1j},\ldots,\epsilon_{nj})$. The risk $MSE(\bm{\widehat\mu},\bm{\mu})$ is then approximated by ${\widehat{MSE}}(\bm{\widehat\mu},\bm{\mu})$, the average of $\sum_{j=1}^q(\hat\mu_j-\mu_j)^2/q$ over all data sets. 

Similarly, we define $MSE(\bm{\widehat{D}},\bm{D})$ and ${\widehat{MSE}}(\bm{\widehat{D}},\bm{D})$ when we are estimating $\bm{D}$.

\subsection{Comparing Different Shrinkage Estimators when $\bm{D}$ is Known} \label{sec:sec4.1} 

In this section, data are generated from model (\ref{eq:LSREeq2}) and it is assumed that $\sigma_1^2,\ldots,\sigma_q^2$ are known. Table \ref{tab:T1} compares $MSE(\bm{\widehat\mu},\bm{\mu})$ for the estimates discussed in \cite{xie2012sure} and \cite{weinstein2018group} with our estimate, denoted \NGam. The estimators of \cite{xie2012sure} defined by their expressions (7.1), (7.2), (7.3), (4.2), (5.1), (6.3), and (6.2) will be called EBMLE.XKB, EBMOM.XKB, JS.XKB, SURE.G.XKB, SURE.M.XKB, SURE.SG.XKB, and SURE.SM.XKB, respectively. \cite{weinstein2018group} developed group-linear and dynamic group-linear algorithms, which are referred to here as GL.WMBZ and DGL.WMBZ, respectively. We also consider Oracle.XKB, which, although not an estimate as described in Section 7 of \cite{xie2012sure}, provides a sensible lower bound on a risk estimator with given parametric form. Our estimator does not belong to this class of estimators because the sample means are not shrunk towards a single value, as discussed in Section \ref{sec:3.1}. 

Examples \ref{exam:exam1}-\ref{exam:exam6} of this section were taken from \cite{xie2012sure} and also used by \cite{weinstein2018group}. We simulated data from model (\ref{eq:LSREeq2}) for different choices of $f_{\mu,\sigma^2}$. The experiment was repeated $1000$ times for each of $q=20,60,100,\ldots,500$. The resulting values of ${\widehat{MSE}}(\bm{\widehat\mu},\bm{\mu})$ are shown in Table \ref{tab:T1} for all $q$ and each of the estimates mentioned above. 

\begin{figure}[hp]
\captionsetup{font=scriptsize}
\centering
\caption{${\widehat{MSE}}(\bm{\widehat\mu},\bm{\mu})$ vs.\ dimension $q$ of normal vector for Examples \ref{exam:exam1}-\ref{exam:exam8} of Section \ref{sec:sec4.1}. The dimension sizes are $q=20,60,\ldots,500$ and results are based on $1000$ replications at each $q$.}\label{fig:examplesknown} 
\includegraphics[width=12cm,height=18cm]{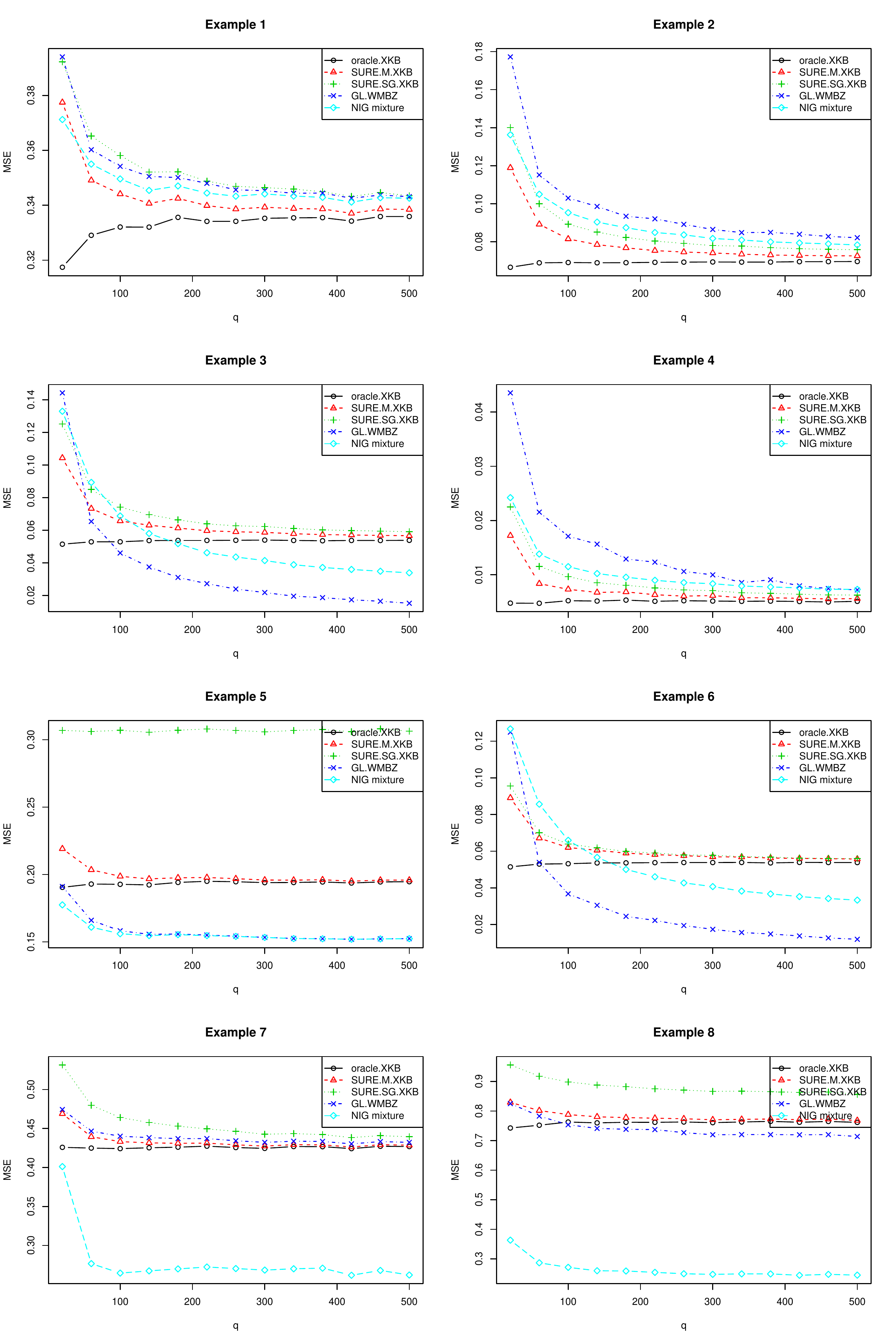}
\end{figure}

\begin{Exa} \label{exam:exam1} 
The density $f_{\mu,\sigma^2}$ is such that $\mu$ and $\sigma^2$ are independent with $\mu \sim N(0,1)$ and $\sigma^2 \sim U(0.1,1)$, where $U(a,b)$ denotes the uniform distribution on the interval $(a,b)$. Here and in Examples 2-5, 7, and 8 we take $\epsilon\sim N(0,1)$. Figure \ref{fig:examplesknown} shows that SURE.M.XKB performs better than SURE.SG.XKB, GL.WMBZ and \NGam\ (the only estimates plotted) since the generated data conform with the parametric form (\ref{eq:heteropost}) upon which SURE.M.XKB is based. Likewise SURE.G.XKB, EBMLE.XKB, and EBMOM.XKB assume that $\bm{\mu}$ has the parametric form (\ref{eq:heteropost}), and hence these estimates outperform the other estimates. Our results (some of which are not given in Figure \ref{fig:examplesknown} or Table \ref{tab:T1}) show that, except for JS.XKB and DGL.WMBZ, all estimated risks converge to the oracle risk. JS.XKB, which applies constant shrinkage for every coordinate, results in an inefficient estimator. Interestingly, even though the distribution of $\sigma^2$ is uniform, the case where group-linear algorithms should perform well because of their use of binning, the \NGam\ estimate outperforms the group-linear algorithms for small $q$. 
\end{Exa}

\begin{Exa} \label{exam:exam2}
The density $f_{\mu,\sigma^2}$ is such that $\mu$ and $\sigma^2$ are independent with $\mu \sim U(0,1)$ and $\sigma^2 \sim U(0.1,1)$. This example is quite similar to Example \ref{exam:exam1}, and shows that the parametric form (\ref{eq:heteropost}) is not necessarily important as long as $\mu$ and $\sigma^2$ are independent. The estimated risks of EBMLE.XKB, EBMOM.XKB and SURE.M.XKB all converge to the risk of Oracle.XKB. Figure \ref{fig:examplesknown} shows that SURE.M.XKB and SURE.SG.XKB perform better than the other two estimates. The fact that the normal-inverse gamma mixture allows for a dependency between $\mu$ and $\sigma^2$ may explain why \NGam\ does not perform as well as the SURE estimates. However, \NGam\ performs better than GL.WMBZ. 
\end{Exa}

\begin{Exa} \label{exam:exam3}
Here the joint distribution of $\mu$ and $\sigma^2$ is singular, with $\sigma^2 \sim U(0.1,1)$ and $\mu=\sigma^2$. Rather than being independent, as in Examples \ref{exam:exam1} and \ref{exam:exam2}, $\mu$ and $\sigma^2$ are highly dependent in this case. Even though the SURE.M.XKB and SURE.SG.XKB risks converge to the Oracle.XKB risk, the Oracle.XKB risk is actually larger than that of GL.WBMZ and \NGam. When $\mu$ and $\sigma^2$ are dependent, SURE estimates tend to perform poorly compared to group-linear algorithms and \NGam. GL.WBMZ is based on clustering $\textrm{log}(\sigma^2)$, and if $\mu$ is a function of $\sigma^2$ then group-linear algorithms will usually cluster the $\mu_j$s correctly, regardless of the distribution of $\mu$. So in this example, group-linear estimates outperform all the other estimates. 
\end{Exa}

\begin{Exa} \label{exam:exam4}
Again the joint distribution of $\mu$ and $\sigma^2$ is singular with $\mu=\sigma^2$, but now $\frac{1}{\sigma^2} \sim \chi^2_{10}$. The risks of SURE.M.XKB and SURE.SG.XKB converge to that of Oracle.XKB as $q$ increases. The \NGam\ estimate performs better than GL.WMBZ for lower values of $q$, but as $q$ increases performance of both of these algorithms improves and approaches that of Oracle.XKB. 
\end{Exa}

\begin{Exa} \label{exam:exam5}
In this example the distribution of $\sigma^2$ is discrete and such that $\sigma^2$ is either $0.1$ or $0.5$, each with probability $1/2$, while $\mu|(\sigma^2=0.1) \sim N(2,0.1)$ and $\mu|(\sigma^2=0.5) \sim N(0,0.5)$. Obviously $\mu$ and $\sigma^2$ are not independent in this case, and there are two distinct groups of data. Both GL.WBMZ and \NGam\ effectively treat the two groups separately, whereas SURE.M.XKB and SURE.SG.XKB shrink all means in the same direction, as does Oracle.XKB. For each $q$, GL.WMBZ and \NGam\ greatly outperform the SURE estimates. 
\end{Exa}

\begin{Exa} \label{exam:exam6}
Here the setting is the same as in Example 3 except that $\epsilon \sim U(-\sqrt{3},\sqrt{3})$. As in Example \ref{exam:exam5}, for any $q$, GL.WMBZ and \NGam\ outperform the SURE estimates and GL.WMBZ performs better than \NGam\ since $\mu$ is a function of $\sigma^2$. 
\end{Exa}

\begin{Exa} \label{exam:exam7}
The density $f_{\mu,\sigma^2}$ is such that $\mu$ and $\sigma^2$ are independent with $\sigma^2 \sim U(0.1,1)$ and $\mu \sim 0.5 N(0,0.1)+ 0.5 N(3,0.1)$. Here the distribution of $\mu$ is bimodal. This is a case where algorithms based on clustering $\sigma^2$ fail, and \NGam\ does very well. SURE estimates shrink all $X_i$ in the same direction, towards 1.5, whereas \NGam\ shrinks $X_j$ towards either $0$ or $3$ after identifying the cluster to which $\mu_j$ is likely to belong. Group-linear estimates end up having the same defect in this case as the SURE estimates. Since clustering is based on $\textrm{log}(\sigma_j^2)$ and $\mu_j$ is independent of $\sigma_j^2$, each group-linear cluster will contain roughly equal numbers of $\mu_j$s from the two components. It follows that the group-linear algorithms will also shrink $X_j$ towards 1.5.
\end{Exa}

\begin{Exa} \label{exam:exam8}
The distribution of $(\mu,\sigma^2)$ is such that $(\mu,\sigma^2) \sim 0.6 N\Gamma^{-1}(2,2,5,2) + 0.4 N\Gamma^{-1}(10,4,3,3)$. In this example the underlying distribution of $(\mu,\sigma^2)$ is a mixture of normal-inverse gammas, and so, as expected, \NGam\ estimate outperforms all the others. As the marginal distribution of $\mu$ is bimodal, SURE and group-linear estimates do not perform well for the same reason as in Example \ref{exam:exam7}.
\end{Exa}

\begin{table}[hp]
\scriptsize
\centering
\captionsetup{font=scriptsize}
\caption[T1]{Averages of $\Mhat$ over all $q=20,60,\ldots,500$ in model (\ref{eq:LSREeq2}) for Examples \ref{exam:exam1}-\ref{exam:exam8} of Section \ref{sec:sec4.1}. For a given $q$, $\Mhat$ is an average over 1000 replications.}
\begin{tabular}{l cccccccc }
\hline
\hline
Different & & & & Example & & & &\\
\cline{2-9}
estimates & \ref{exam:exam1} & \ref{exam:exam2} & \ref{exam:exam3} & \ref{exam:exam4} & \ref{exam:exam5} & \ref{exam:exam6} & \ref{exam:exam7} & \ref{exam:exam8}\\
\hline
Sample Statistics 	&	0.5504	&	0.5496	&	0.5506	&	0.1248	&	0.3008	&	0.5502	&	0.5506	&	0.8976	\\
EBMLE.XKB 	&	0.3410	&	0.0762	&	0.0833	&	0.0071	&	0.2524	&	0.0814	&	0.4311	&	0.8448	\\
EBMOM.XKB 	&	0.3412	&	0.0832	&	0.0906	&	0.0086	&	0.2467	&	0.0822	&	0.4313	&	0.8423	\\
JS.XKB 	&	0.3675	&	0.0837	&	0.0885	&	0.0075	&	0.2616	&	0.085	&	0.4523	&	0.8563	\\
Oracle.XKB 	&	0.3328	&	0.0691	&	0.0535	&	0.0051	&	0.1936	&	0.0535	&	0.4258	&	0.7602	\\
SURE.G.XKB 	&	0.3424	&	0.0792	&	0.0645	&	0.0072	&	0.2365	&	0.0613	&	0.4327	&	0.8393	\\
SURE.M.XKB 	&	0.3433	&	0.0795	&	0.0639	&	0.0072	&	0.1988	&	0.0608	&	0.4334	&	0.7811	\\
SURE.SG.XKB 	&	0.3526	&	0.086	&	0.0699	&	0.0088	&	0.3068	&	0.0621	&	0.4561	&	0.8824	\\
SURE.SM.XKB 	&	0.3557	&	0.0877	&	0.0698	&	0.0091	&	0.1877	&	0.0628	&	0.4569	&	0.6829	\\
GL.WMBZ	&	0.3512	&	0.098	&	0.0373	&	0.0141	&	0.1578	&	0.0306	&	0.4387	&	0.7401	\\
GL.SURE.WMBZ	&	0.3534	&	0.0974	&	0.0473	&	0.0127	&	0.1578	&	0.0368	&	0.4415	&	0.7249	\\
DGL.WMBZ	&	0.3714	&	0.1155	&	0.1044	&	0.0158	&	0.2496	&	0.0937	&	0.4523	&	0.8525	\\
$N\Gamma^{-1}$ mixture	&	0.3471	&	0.0894	&	0.0548	&	0.0102	&	0.1560	&	0.0532	&	0.2787	&	0.2639	\\
\hline
\hline\label{tab:T1}
\end{tabular}
\end{table}

\subsection{Comparing Different Shrinkage Estimators when $\bm{D}$ is Unknown} \label{sec:sec4.2} 
 
Tables \ref{tab:T21} and \ref{tab:T22} compare the different estimates discussed in \cite{xie2012sure}, \cite{weinstein2018group} and \cite{jing2016sure}. The estimate referred to as SURE.M.Double can be found in (11)-(12) of \cite{jing2016sure}. Although \cite{jing2016sure} discussed a few different double shrinkage algorithms, we have found the performance of those algorithms to be very similar to each other, and therefore report results only for the algorithm in expression (16) of \cite{jing2016sure}, which we refer to as SURE.M.Double. As \cite{xie2012sure} and \cite{weinstein2018group} assumed that $\sigma_1^2,\ldots,\sigma_q^2$ were known, we do as they suggested and replace $\sigma_j^2$ by $S_{\cdot j}^2$ when implementing their algorithms.

\begin{figure}[hp]
\captionsetup{font=scriptsize}
\centering
\caption{$\Mhat$ vs.\ dimension $q$ of normal vector for Example \ref{exam:exam9}-\ref{exam:exam14} of Section \ref{sec:sec4.2}. The dimension sizes are $q=20,60,\ldots,500$ and results are based on $1000$ replications at each $q$.}\label{fig:examplesunknown}
\includegraphics[width=12cm,height=15cm]{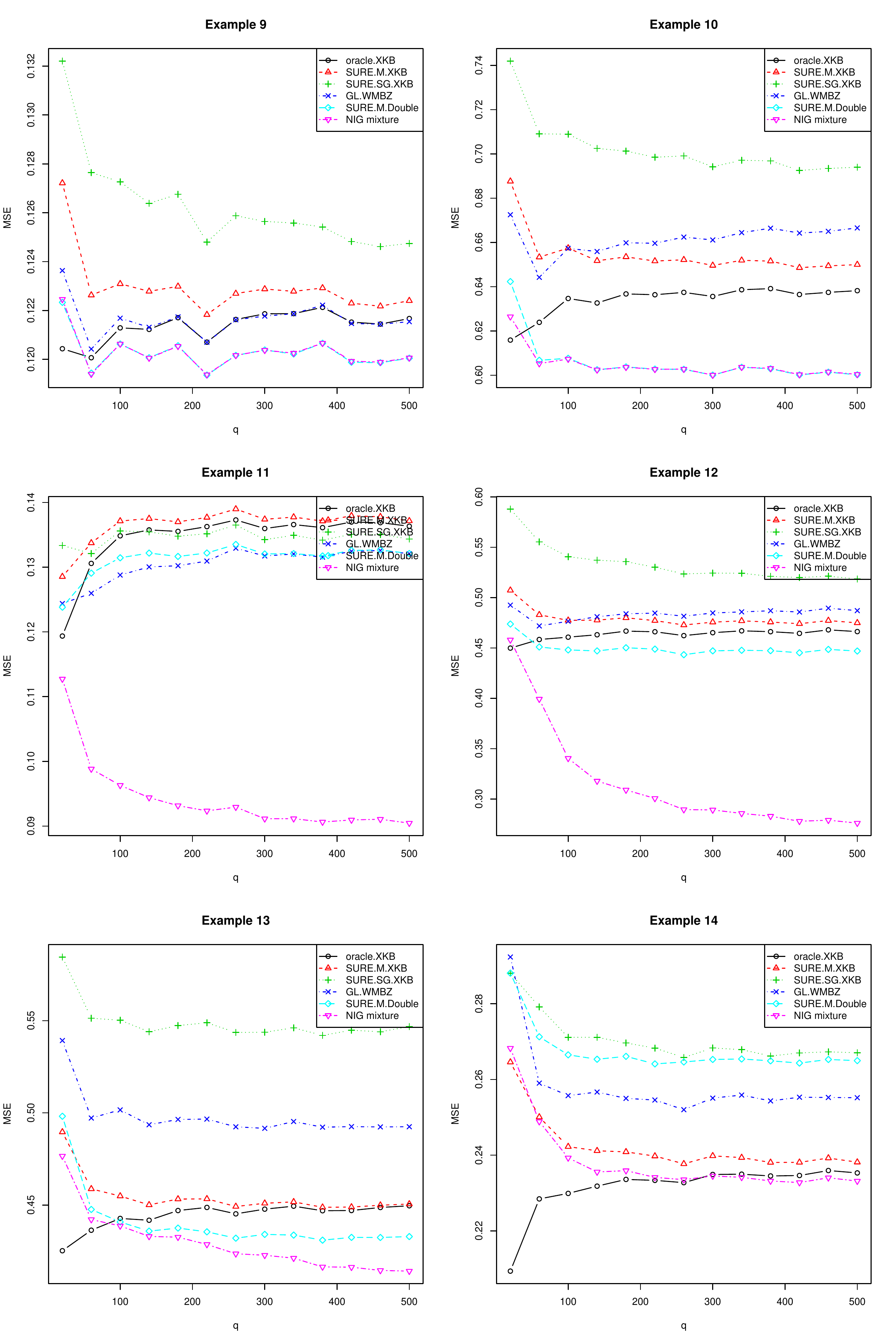}
\end{figure}

\begin{figure}[hp]
\captionsetup{font=scriptsize}
\centering
\caption{$\MhatD$ vs.\ dimension $q$ of normal vector for Examples \ref{exam:exam9}-\ref{exam:exam14} of Section \ref{sec:sec4.2}. The dimension sizes are $q=20,60,\ldots,500$ and results are based on $1000$ replications at each $q$.}\label{fig:s-examplesunknown}
\includegraphics[width=12cm,height=15cm]{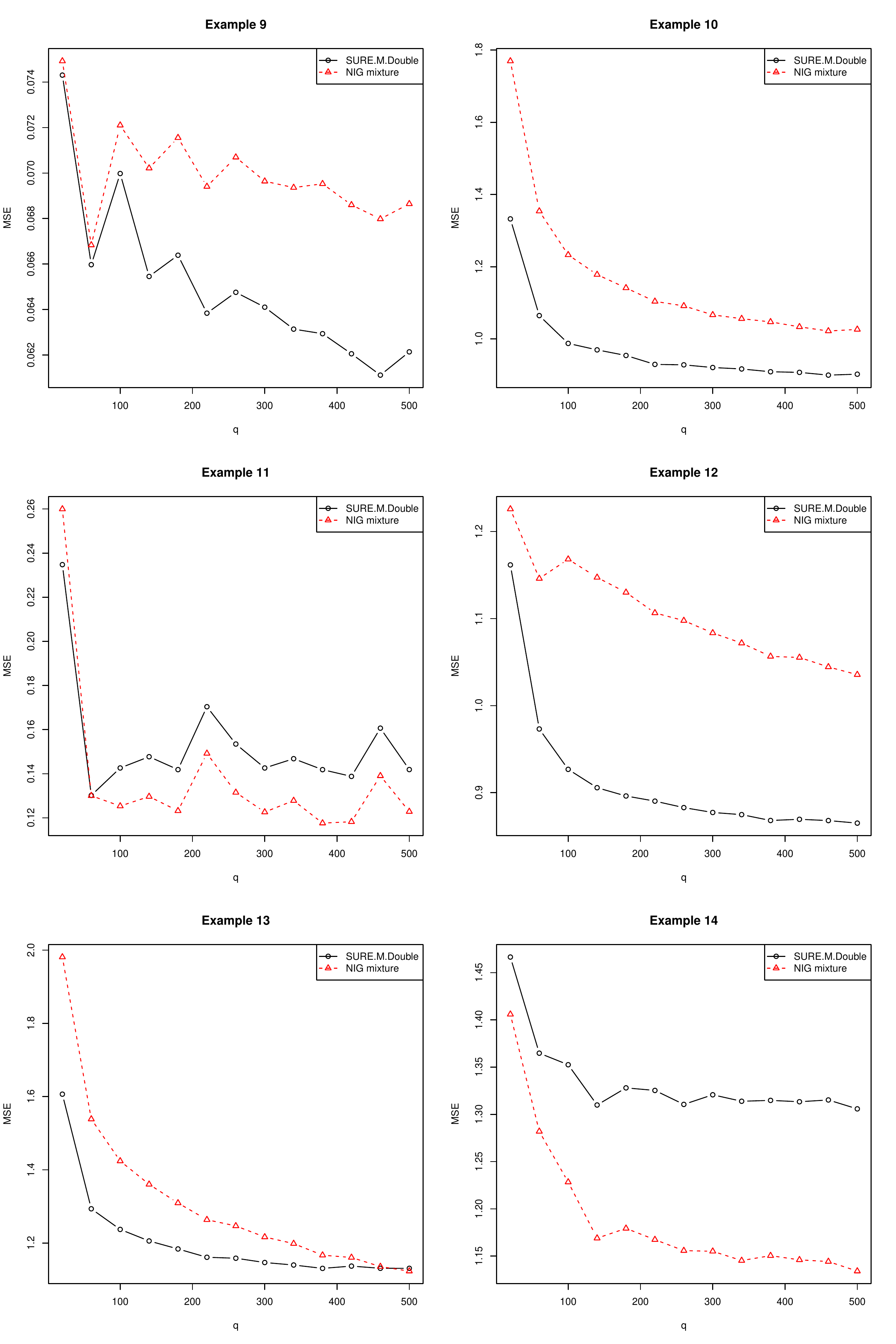}
\end{figure}

We simulated data from model (\ref{eq:LSREeq1}) for different choices of $f_{\mu,\sigma^2}$. In all the examples of this section $\epsilon \sim N(0,1)$. For each $(\mu_j,\sigma_j^2)$ pair there are $n=4$ replications. We only observe $X_{ij}$, for $i=1,\ldots,n$, $j=1,\ldots,q$, and not $\sigma_1^2,\ldots,\sigma_q^2$. We repeat the experiment 1000 times for each $q$, and $\Mhat$ and $\MhatD$ were determined. Tables \ref{tab:T21} and \ref{tab:T22} provide estimated risks averaged over all $q$, and Figure \ref{fig:examplesunknown} shows how our estimate compares with the two SURE estimates discussed in \cite{xie2012sure} and with the group-linear algorithms discussed in \cite{weinstein2018group}. Figure \ref{fig:s-examplesunknown} shows how our estimate of estimating $\bm{D}$ compares with the SURE.M.Double discussed in \cite{jing2016sure}. 

\begin{Exa} \label{exam:exam9}
The density $f_{\mu,\sigma^2}$ is such that $\mu$ and $\sigma^2$ are independent with $\mu \sim N(0,3)$ and $\sigma^2 \sim IG(5,2)$. Figure \ref{fig:examplesunknown} shows that our estimate outperforms the other three when estimating $\mu_j$. Table \ref{tab:T21} shows that \NGam\ performs similarly to the double shrinkage algorithms discussed in \cite{jing2016sure}. As the latter algorithms and \NGam\ are based on the normal-inverse gamma distribution, and the $(\mu_j,\sigma_j^2)$ distribution in this case is normal-inverse gamma, it is not surprising that these estimates outperform the others here. Table \ref{tab:T22} and Figure \ref{fig:s-examplesunknown} show that the SURE.M.Double estimate slightly outperforms \NGam\ in estimating $\bm{D}$. 
\end{Exa}

\begin{Exa} \label{exam:exam10}
The density $f_{\mu,\sigma^2}$ is such that $\mu$ and $\sigma^2$ are independent with $\mu \sim N(0,3)$ and $\sigma^2 \sim G(9,3)$. This case is similar to Example \ref{exam:exam7}, and likewise the results are similar.
\end{Exa}

\begin{Exa} \label{exam:exam11}
Here $(\mu,\sigma^2) \sim 0.95 N{\Gamma}^{-1}(2,2,5,2)+0.05 N{\Gamma}^{-1}(10,4,3,3)$, the same mixture distribution considered in Example \ref{exam:exam8}. This is a case where $\mu$ and $\sigma^2$ are dependent and their distribution is bimodal. Our algorithm outperforms all other estimates in terms of both $\bm{\mu}$ and $\bm{D}$ estimation, as seen in Figures \ref{fig:examplesunknown}-\ref{fig:s-examplesunknown} and Tables \ref{tab:T21}-\ref{tab:T22}.
\end{Exa}

\begin{Exa} \label{exam:exam12}
In this case $\mu$ and $\sigma^2$ are independent with $\mu \sim 0.5 U(1,2)+0.5 U(4,5)$ and $\frac{\sigma^2}{n} \sim U(0.1,1)$. This is a case where $\mu$ and $\sigma^2$ are independent and have a bimodal distribution. As in Example \ref{exam:exam11}, the \NGam\ estimate outperforms all other estimates with respect to estimating $\mu$. However, presumably because the distribution of $\sigma^2$ is unimodal, SURE estimates do better in terms of estimating $\sigma^2$. 
\end{Exa}

\begin{Exa} \label{exam:exam13}
The distribution of $(\mu,\sigma^2)$ is such that $\mu \sim N(3,1^2)$ and $\sigma^2|\mu \sim U(\textrm{max}(\mu-1,0.1),\textrm{max}(\mu+1,1))$. Here, $\mu$ and $\sigma^2$ are dependent, which is a case where SURE estimates do not perform well. The \NGam\ estimate outperforms the other estimates in terms of $\Mhat$ and in terms of $\MhatD$ for larger $q$. 
\end{Exa}

\begin{Exa} \label{exam:exam14}
The distribution of $(\mu,\sigma^2)$ is such that $\mu \sim N(3,1^2)$ and $\sigma^2|\mu \sim \textrm{max}(N(\frac{|\mu|}{3},$ ${(\frac{|\mu|}{3}+1)}^2),0.1)$. Again, since $\mu$ and $\sigma^2$ are dependent, the SURE estimates do not perform well. The group-linear algorithms lose efficiency as $\sigma_j^2$ is replaced by $S_{\cdot j}^2$, and the \NGam\ estimate outperforms all other estimates in terms of both $\Mhat$ and $\MhatD$.
\end{Exa}

\begin{table}[hp]
\scriptsize
\centering
\captionsetup{font=scriptsize}
\caption[T21]{
Averages of $\Mhat$ over all $q=20,60,\ldots,500$ in model (\ref{eq:LSREeq1}) for Examples \ref{exam:exam9}-\ref{exam:exam14} of Section \ref{sec:sec4.2}. For a given $q$, $\Mhat$ is an average over 1000 replications.}
\begin{tabular}{l cccccc }
\hline
\hline
Different & & & Example & & & \\
\cline{2-7}
estimates & \ref{exam:exam9}& \ref{exam:exam10}& \ref{exam:exam11}& \ref{exam:exam12}& \ref{exam:exam13}& \ref{exam:exam14}\\
\hline
Sample Statistics 	&	0.1247	&	0.7484	&	0.1376	&	0.5520	&	0.7525	&	0.3570	\\
EBMLE.XKB 	&	0.1217	&	0.6369	&	0.1795	&	0.4681	&	0.4925	&	0.2432	\\
EBMOM.XKB 	&	0.1217	&	0.6381	&	0.1710	&	0.4690	&	0.4881	&	0.2430	\\
JS.XKB 	&	0.1222	&	0.6637	&	0.1328	&	0.5023	&	0.5997	&	0.3416	\\
Oracle.XKB 	&	0.1214	&	0.6350	&	0.1362	&	0.4670	&	0.4491	&	0.2342	\\
SURE.G.XKB 	&	0.1223	&	0.6479	&	0.1373	&	0.4765	&	0.4704	&	0.2428	\\
SURE.M.XKB 	&	0.1224	&	0.6483	&	0.1371	&	0.4769	&	0.4513	&	0.2384	\\
SURE.SG.XKB 	&	0.1249	&	0.6927	&	0.1343	&	0.5215	&	0.5461	&	0.2662	\\
SURE.SM.XKB 	&	0.1252	&	0.6954	&	0.1343	&	0.5228	&	0.5289	&	0.2638	\\
GL.WMBZ	&	0.1216	&	0.6644	&	0.1317	&	0.4882	&	0.4958	&	0.2544	\\
GL.SURE.WMBZ	&	0.1220	&	0.6720	&	0.1310	&	0.4965	&	0.5020	&	0.2589	\\
DGL.WMBZ	&	0.1200	&	0.6045	&	0.1312	&	0.4538	&	0.4430	&	0.2679	\\
SURE.M.Double	&	0.1199	&	0.5995	&	0.1319	&	0.4493	&	0.4340	&	0.2649	\\
$N\Gamma^{-1}$ mixture	&	0.1198	&	0.5995	&	0.0911	&	0.2849	&	0.4176	&	0.2333	\\
\hline
\hline\label{tab:T21}
\end{tabular}
\end{table}

\begin{table}[hp]
\scriptsize
\centering
\captionsetup{font=scriptsize}
\caption[T22]{Averages of $\MhatD$ over all $q=20,60,\ldots,500$ in model (\ref{eq:LSREeq1}) for Examples \ref{exam:exam9}-\ref{exam:exam14} of Section \ref{sec:sec4.2}. For a given $q$, $\MhatD$ is an average
 over 1000 replications.}
\begin{tabular}{l cccccc }
\hline
\hline
Different & & & Example & & & \\
\cline{2-7}
estimates & \ref{exam:exam9}& \ref{exam:exam10}& \ref{exam:exam11}& \ref{exam:exam12}& \ref{exam:exam13}& \ref{exam:exam14}\\
\hline
Sample Statistics 	&	0.2206	&	6.6980	&	0.3836	&	3.9425	&	1.1918	&	2.9284	\\
SURE.M.Double 	&	0.0626	&	0.9160	&	0.1548	&	0.8692	&	0.2873	&	1.3108	\\
$N\Gamma^{-1}$	&	0.0689	&	1.0065	&	0.1283	&	0.9630	&	0.3072	&	1.0025	\\
\hline
\hline \label{tab:T22}
\end{tabular}
\end{table}

\begin{figure}[hp]
\captionsetup{font=scriptsize}
\centering
\caption{Box plot of estimated $\pi_{(8)}$, the $3^{rd}$ highest probability for $q=20,60,\ldots,500$ in model (\ref{eq:LSREeq1}) for Example \ref{exam:exam11} of Section \ref{sec:sec4.2}. For a given $q$, each boxplot is drawn using 100 simulations.}\label{fig:sensitivity_gamma4}
\includegraphics[width=12cm,height=10cm]{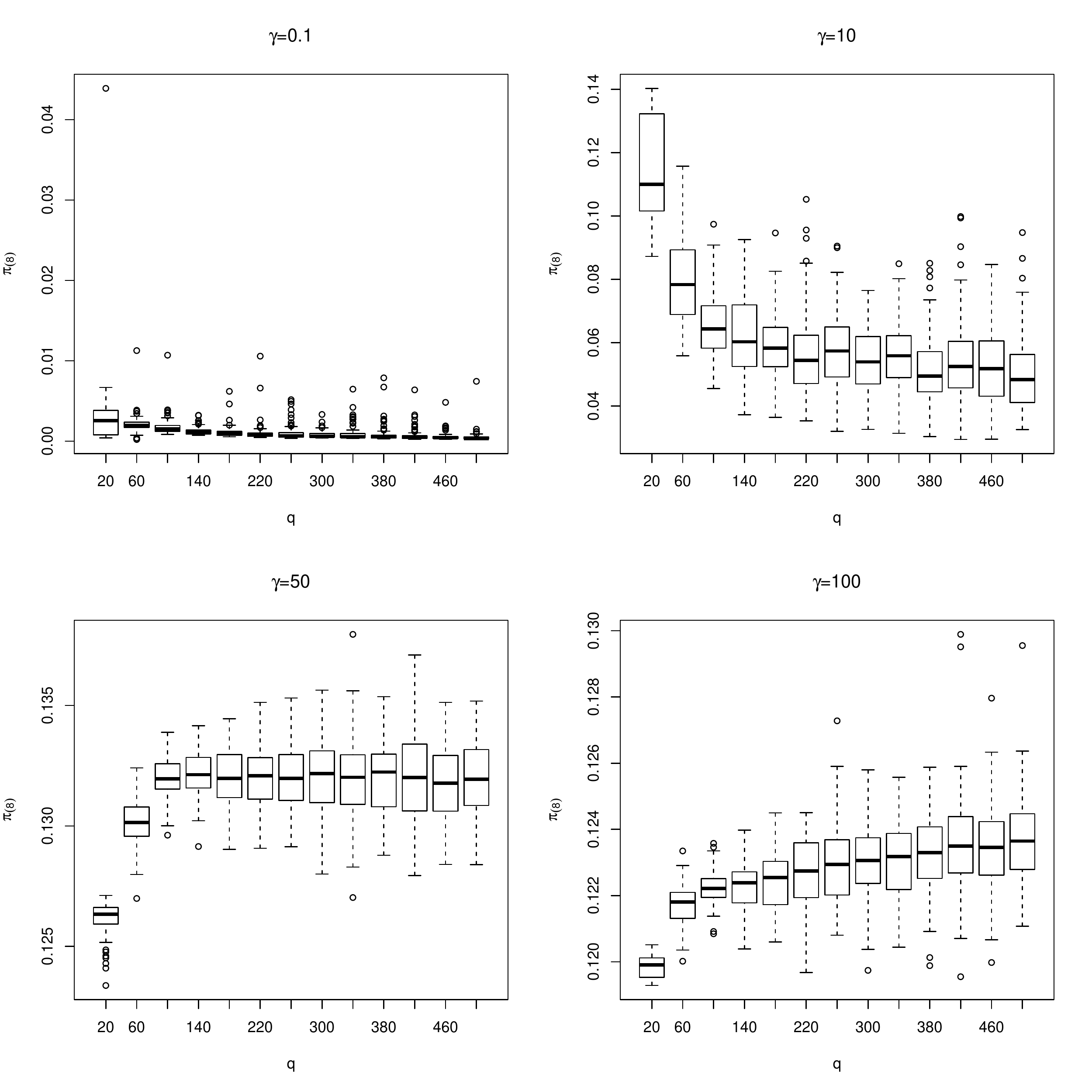}
\end{figure}

\begin{figure}[hp]
\captionsetup{font=scriptsize}
\centering
\caption{Box plot of estimated $\pi_{(9)}$, the $2^{nd}$ highest probability for $q=20,60,\ldots,500$ in model (\ref{eq:LSREeq1}) for Example \ref{exam:exam11} of Section \ref{sec:sec4.2}. The true value of this parameter is 0.05. For a given $q$, each boxplot is drawn using 100 simulations.}\label{fig:sensitivity_gamma3}
\includegraphics[width=12cm,height=10cm]{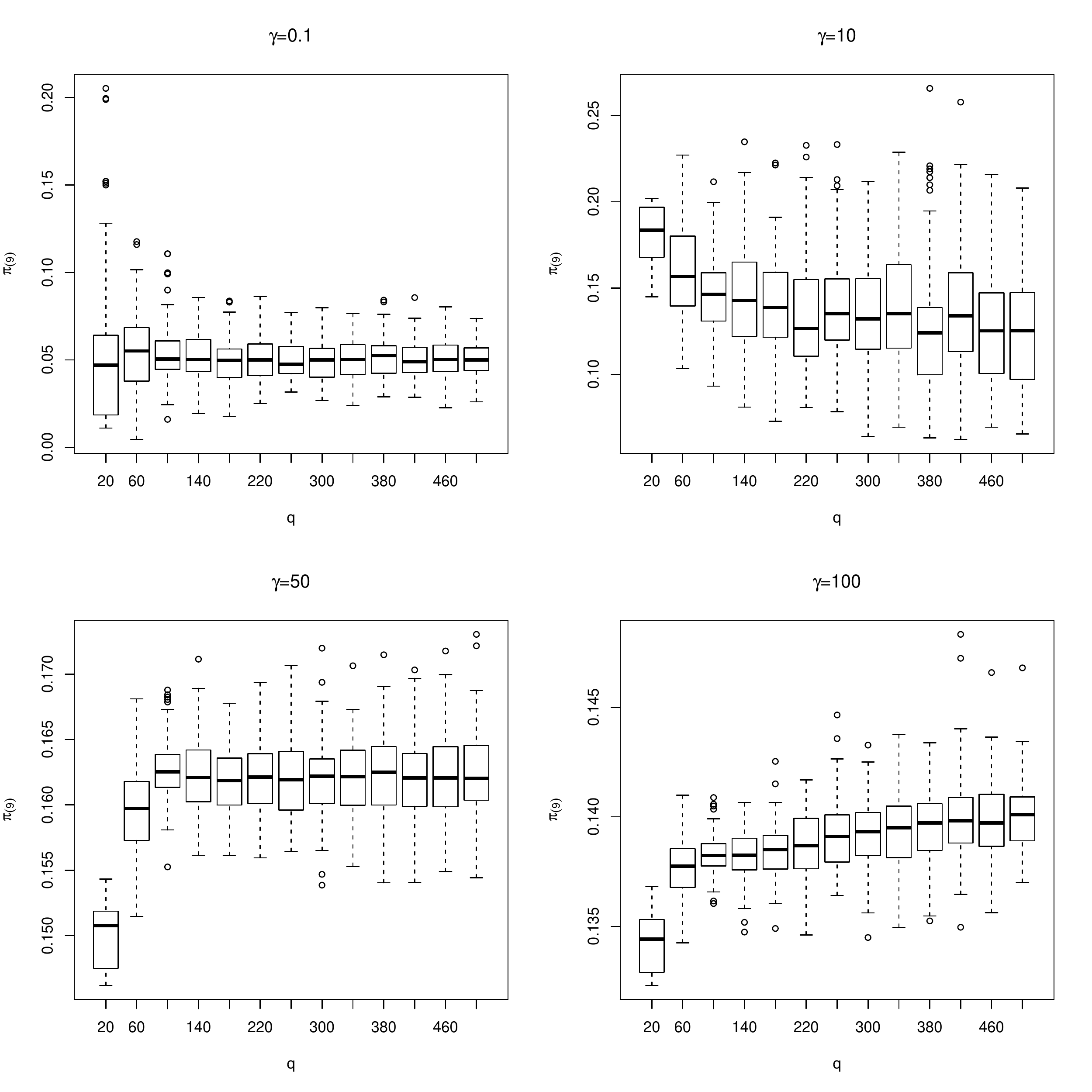}
\end{figure}

We experimented with several choices for $\gamma$, the DPMM
concentration parameter. In Example \ref{exam:exam11} we took
  $\gamma$ to be 0.1, 10, 50, 100 
  with $k=10$. Since the true distribution
  is bimodal, ideally the DP process should have only two active
  components. Table 
  \ref{tab:T5} shows changing the value of $\gamma$ has very little
  effect on the mean squared error of either $\hat{\bm{\mu}}$ or $\hat{\bm{D}}$. On the other hand Figures \ref{fig:sensitivity_gamma4} and \ref{fig:sensitivity_gamma3} show that a lower value of $\gamma$ does a better job of selecting the number of clusters. The higher values of $\gamma$ over-select the number of active components but at least two of the active components have
  very similar \NGam\ parameters, implying that the shrinkage direction and
  factor for each $\mu_j$ changes little. The result is almost no
  change in mean squared error as long as $k < q$.
  For Figure \ref{fig:sensitivity_gamma4} and \ref{fig:sensitivity_gamma3}, we need to estimate $\bm{\pi}$, which we are calculating by averaging over all MCMC iteration. Due to the non-identifiability arising from permutations of the labels in the mixture representation, we sort $\bm{\pi}$ in every MCMC iteration, and then average results estimate the $2^{nd}$ and $3^{rd}$ highest probabilities. Presumably sorting $\bm{\pi}$ in every iteration will mitigate the label switching problem as in the true $f_{\mu,\sigma^2}$ two mixing probabilities are very different from each other.

\begin{table}[hp]
\scriptsize
\centering
\captionsetup{font=scriptsize}
\caption[T22]{Averages of measures over all $q=20,60,\ldots,500$ in model (\ref{eq:LSREeq1}) for Example \ref{exam:exam11} of Section \ref{sec:sec4.2}. For a given $q$, each measure is an average
 over 100 replications.}
\begin{tabular}{l cccc }
\hline
\hline
Different & & & & \\
\cline{2-5}
measures & $\gamma=0.1$& $\gamma=10$& $\gamma=50$& $\gamma=100$\\
\hline
$\Mhat$ 	&	0.0937	&	0.0954	&	0.0965	&	0.0966\\
$\MhatD$	&	0.1863	&	0.2063	&	0.2081	&	0.2059\\
\hline
\hline \label{tab:T5}
\end{tabular}
\end{table}

\section{Real data example when ${\bf D}$ is known} \label{sec:sec5}

In this section, we consider a baseball data example as a test case for our \NGam\ mixture estimate. This data set has been used in the articles of \cite{brown2008season}, \cite{xie2012sure}, \cite{jing2016sure}, and \cite{weinstein2018group}. The data consist of the entire season batting records for all major league baseball players in the 2005 season. The goal is to estimate batting averages of individual players in the second half of the season by observing only the first half averages. Following the other articles, only players with at least 11 at-bats in the first half of the season were considered in the estimation process, and only players with at least 11 at-bats in each of the two halves of the season were considered in the validation process. 

Let $H_{ij}$ denote the number of hits and $N_{ij}$ the number of at-bats for player $j$ in period $i$. The subscript $i$ indicates either the first or second half of the season. The quantity $p_j$ denotes the probability of a hit for player $j$. Then we assume that 
\begin{align*}
H_{ij} \sim \textrm{Bin}(N_{ij},p_j), \quad \textrm{for} \quad i=1,2,\quad j=1,\ldots,q,
\end{align*}
where $\textrm{Bin}(n,p)$ denotes a binomial distribution with
  number of trials $n$ and probability of success $p$. Without doing
any variance-stabilizing transformation, \cite{jing2016sure} worked
with the sample proportion $X_{1j}=H_{1j}/N_{1j}$ and the estimated
variance, $S_{1j}^2=(X_{1j}(1-X_{1j}))/N_{1j}$, of $X_{1j}$. However,
this contradicts their initial assumption that $X_{1j}$ and $S_{1j}^2$
are independently distributed. Also, without the transformation there
is no reason to believe that $X_{1j}$ is normally distributed and
$S_{1j}^2$ follows a chi-square distribution. So, we will follow the
transformation of \cite{brown2008season}, which was also used in
\cite{xie2012sure} and \cite{weinstein2018group}, and define  
\begin{align*}
X_{ij} = \textrm{arcsin} \sqrt{\frac{H_{ij}+0.25}{N_{ij}+0.5}},
\end{align*}
resulting in
\begin{align*}
X_{ij} \dot\sim N(\mu_j, \sigma_{ij}^2), \quad \mu_j=\textrm{arcsin}\left(\sqrt p_j\right), \quad \sigma_{ij}^2={(4N_{ij})}^{-1}. 
\end{align*}
The measure of error that was used in all these papers, denoted TSE, is used to compare different estimates: 
\begin{align*}
TSE(\bm{\hat{\mu}})= \frac{\sum_{j} {(X_{2j}-\hat{\mu}_j)}^2 - \sum_{j} {(4N_{2j})}^{-1}}{\sum_{j} {(X_{2j}-X_{1j})}^2 - \sum_{j} {(4N_{2j})}^{-1}}. 
\end{align*}

The transformed data are consistent with model (\ref{eq:LSREeq2}) as all $\sigma_j^2$ are known. The MCMC algorithm described in Section \ref{sec:sec3} is modified here by simply removing the step of updating $\sigma_j^2$. Table \ref{tab:T3} is the table from \cite{weinstein2018group} with our estimate added in the bottom row.

\begin{table}[H]
\scriptsize
\centering
\captionsetup{font=scriptsize}
\caption[T3]{Average Prediction error for transformed batting averages. $TSE(\bm{\hat{\mu}})$ was computed for the entire data set, and separately for pitchers and non-pitchers.}
\begin{tabular}{l ccc }
\hline
\hline
Different & & Data sets &\\
\cline{2-4}
estimates & All& Pitchers & Non-pitchers\\
\hline
Naive	&	1	&	1	&	1	\\
Grand mean	&	0.852	&	0.127	&	0.378	\\
Nonparametric EB	&	0.508	&	0.212	&	0.372	\\
Binomial mixture	&	0.588	&	0.156	&	0.314	\\
Weighted Least Squares	&	1.07	&	0.127	&	0.468	\\
Weighted nonparametric MLE	&	0.306	&	0.173	&	0.326	\\
Weighted Least Squares (AB)	&	0.537	&	0.087	&	0.29	\\
Weighted nonparametric MLE (AB)	&	0.301	&	0.141	&	0.261	\\
JS.XKB	&	0.535	&	0.165	&	0.348	\\
SURE.M.XKB	&	0.421	&	0.123	&	0.289	\\
SURE.SG.XKB	&	0.408	&	0.091	&	0.261	\\
GL.WMBZ	&	0.302	&	0.178	&	0.325	\\
DGL.WMBZ	&	0.288	&	0.168	&	0.349	\\
$N\Gamma^{-1}$ mixture	&	0.361	&	0.161	&	0.292\\
\hline
\hline\label{tab:T3}
\end{tabular}
\end{table}

The naive estimator simply uses $X_{1j}$ to predict $X_{2j}$ and has TSE equal to 1. The grand mean uses the average of all $X_{1j}$ to predict any $X_{2j}$. The nonparametric EB estimate of \cite{brown2009nonparametric}, the binomial mixture of \cite{muralidharan2010empirical}, the weighted least squares estimator, the weighted least squares estimator (AB) (with number of at-bats as covariate), the weighted nonparametric MLE and the weighted nonparametric MLE (AB) (with number of at-bats as covariate) of \cite{jiang2009general} are also included in Table \ref{tab:T3}. 

\cite{weinstein2018group} presented an analysis under permutations, where each permutation is the order in which successful hits appear throughout the entire season. For each player they draw the number of hits in $N_{1j}$ at-bats from a hypergeometric distribution, $HG(N_{1j}+N_{2j}, H_{1j}+H_{2j}, N_{1j})$. We compare our estimate with several other estimates with respect to $1000$ different permutations of the baseball data and average TSE.

As discussed in \cite{weinstein2018group}, group-linear algorithms tend to perform well compared to SURE estimates as $\mu_j$ and $\sigma_{1j}^2$ are not independent, owing to the fact that players with higher batting averages tend to play more. Also, non-pitchers tend to have higher batting averages than pitchers, so it is possible that the underlying density of $\mu$ is bimodal. This may be the reason that empirical Bayes estimators that assume a normal-normal model tend to perform poorly. group-linear estimates outperform the other estimates because they can accommodate these features exhibited by the baseball data. SURE estimates work well when we analyze the pitchers and non-pitchers separately. Table \ref{tab:T3} shows that, in the combined data, the \NGam\ estimate does not perform as well as group-linear algorithms, but it performs better than SURE estimates. However, when the pitchers and non-pitchers are considered separately, \NGam\ performs better than the group-linear algorithms. In both the original data and the permuted data, \NGam\ performs better than the group-linear algorithms for both pitchers and non-pitchers. When pitchers and non-pitchers are combined, group-linear estimates outperform all other estimates in both the original and permuted data. This is reasonable as the association between $\mu$ and $\sigma^2$ is weaker when the data are separated into smaller groups, and group-linear algorithms work well in the presence of strong association. In contrast, the \NGam\ estimate works reasonably well $\mu$ and $\sigma^2$ are either strongly or weakly dependent. 

\begin{table}[H]
\scriptsize
\centering
\captionsetup{font=scriptsize}
\caption[T4]{Average Prediction error for 1000 permutations of transformed batting averages data. Average $TSE(\bm{\hat{\mu}})$ was computed for the entire data set, and separately for pitchers and non-pitchers.}
\begin{tabular}{l ccc }
\hline
\hline
Different & & Data sets &\\
\cline{2-4}
estimates & All& Pitchers & Non-pitchers\\
\hline
Grand mean	&	0.9222	&	0.3127	&	0.2951	\\
James-Stein	&	0.5465	&	0.2490	&	0.2304	\\
SURE.M.XKB	&	0.4852	&	0.2227	&	0.2602	\\
SURE.SG.XKB	&	0.4693	&	0.1759	&	0.2148	\\
GL.WMBZ(bins = $q^{1/3}$)	&	0.2798	&	0.2438	&	0.1731	\\
GL.SURE.WMBZ	&	0.3032	&	0.2838	&	0.1949	\\
DGL.WMBZ	&	0.4751	&	0.2193	&	0.2250	\\
$N\Gamma^{-1}$ mixture	&	0.3535	&	0.2377	&	0.1698	\\
\hline
\hline\label{tab:T4}
\end{tabular}
\end{table}

\section{Real data example when ${\bf D}$ is unknown} \label{sec:sec6}

In this section, we will apply the \NGam\ estimate and other estimators to the prostate data from the book of \cite{efron2012large}. The data can be downloaded from the book website \url{https://statweb.stanford.edu/~ckirby/brad/LSI/datasets-and-programs/data/}. The pros- tate data consist of genetic expression levels for $q=6033$ genes obtained from 102 men, 50 normal control and 52 prostate cancer patients. We only use the control data, which means that we have a $50 \times 6033$ matrix. Here $X_{ij}$ denotes the expression level for gene $j$ of patient $i$, $i=1,\ldots,50$, $j=1,\ldots,6033$. Since $50$ is a relatively large number, we will assume that the control group constitutes the population of interest, in which case $$\mu_j=\frac{1}{50} \sum_{i=1}^{50} X_{ij}\quad\textrm{and}\quad\sigma_j^2=\frac{1}{50} \sum_{i=1}^{50} {(X_{ij}-\mu_j)}^2,\quad j=1,\ldots,6033.$$ As a test of the various estimates, we randomly select three subjects from the control group and use their data to estimate $\mu_j$ and $\sigma_j^2$. 

\begin{figure}[hp]
\captionsetup{font=scriptsize}
\centering
\caption{Scatterplots for prostate data. The upper left plot is $S_{\cdot j}^2$ vs.~$\bar{X}_{\cdot j}$ for columns $6$, $30$ and $31$ of the data matrix, the upper right plot is $\sigma_j^2$ vs.~$\mu_j$ and the lower left plot is ${\hat{\sigma}_{j,DPMM}}^{2}$ vs.~${\hat{\mu}_j}^{DPMM}$ based on columns $6$, $30$ and $31$.}\label{fig:prostate1}
\includegraphics[width=12cm,height=10cm]{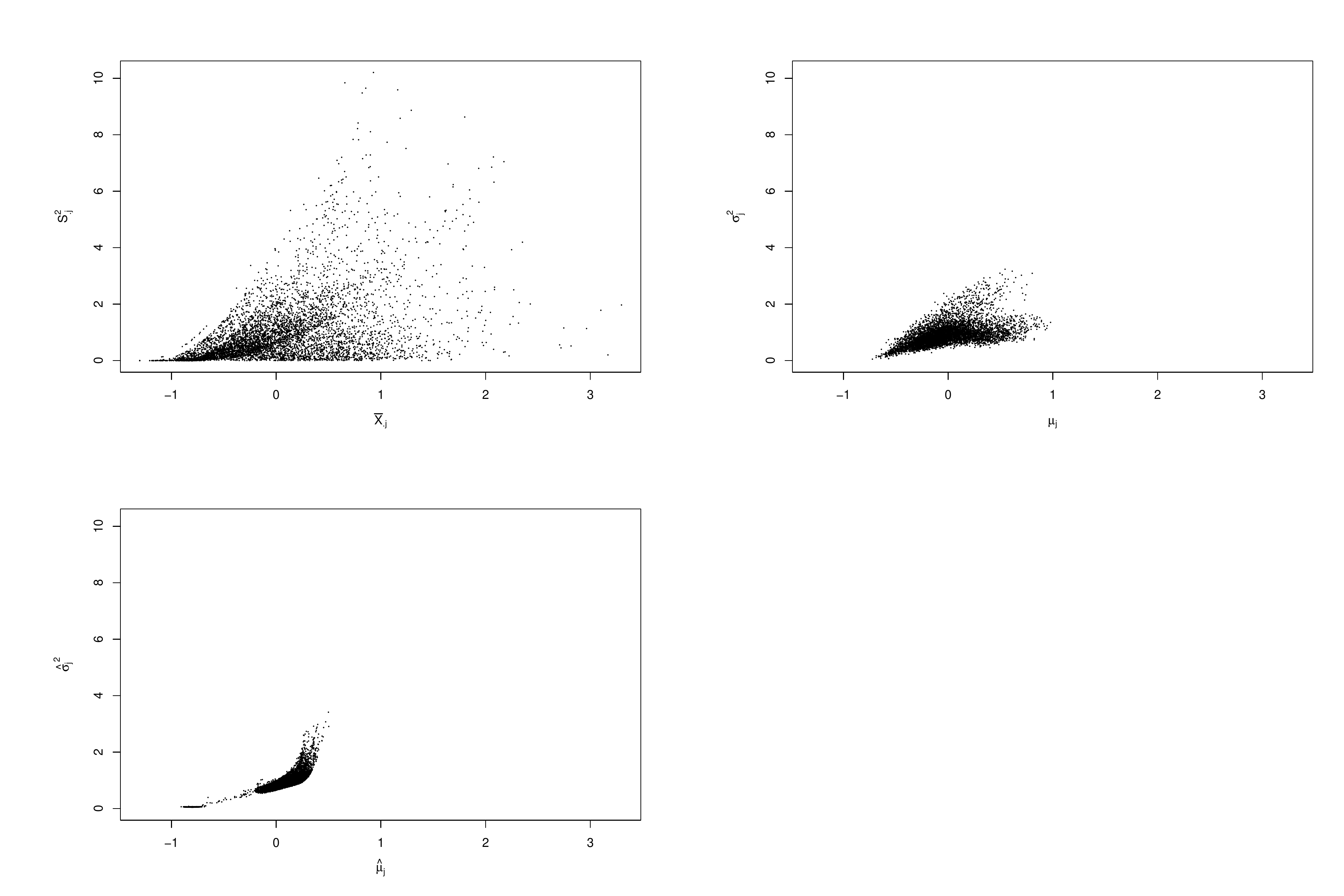}
\end{figure}

\begin{figure}[hp]
\captionsetup{font=scriptsize}
\centering
\caption{Scatterplots for prostate data based 3 columns $6$, $30$ and $31$ of the data matrix. The upper left plot is $\bar{X}_{\cdot j}$ vs.~$\mu_j$, the upper right plot is ${\hat{\mu}_j}^{DPMM}$ vs.~$\mu_j$ based on columns $6$, $30$ and $31$. The lower left plot is $\bar{S}_{\cdot j}^2$ vs.~$\sigma_j^2$, the lower right plot is ${\hat{\sigma}_{j,DPMM}}^{2}$ vs.~$\sigma_j^2$ based on columns $6$, $30$ and $31$.}\label{fig:prostate2}
\includegraphics[width=12cm,height=10cm]{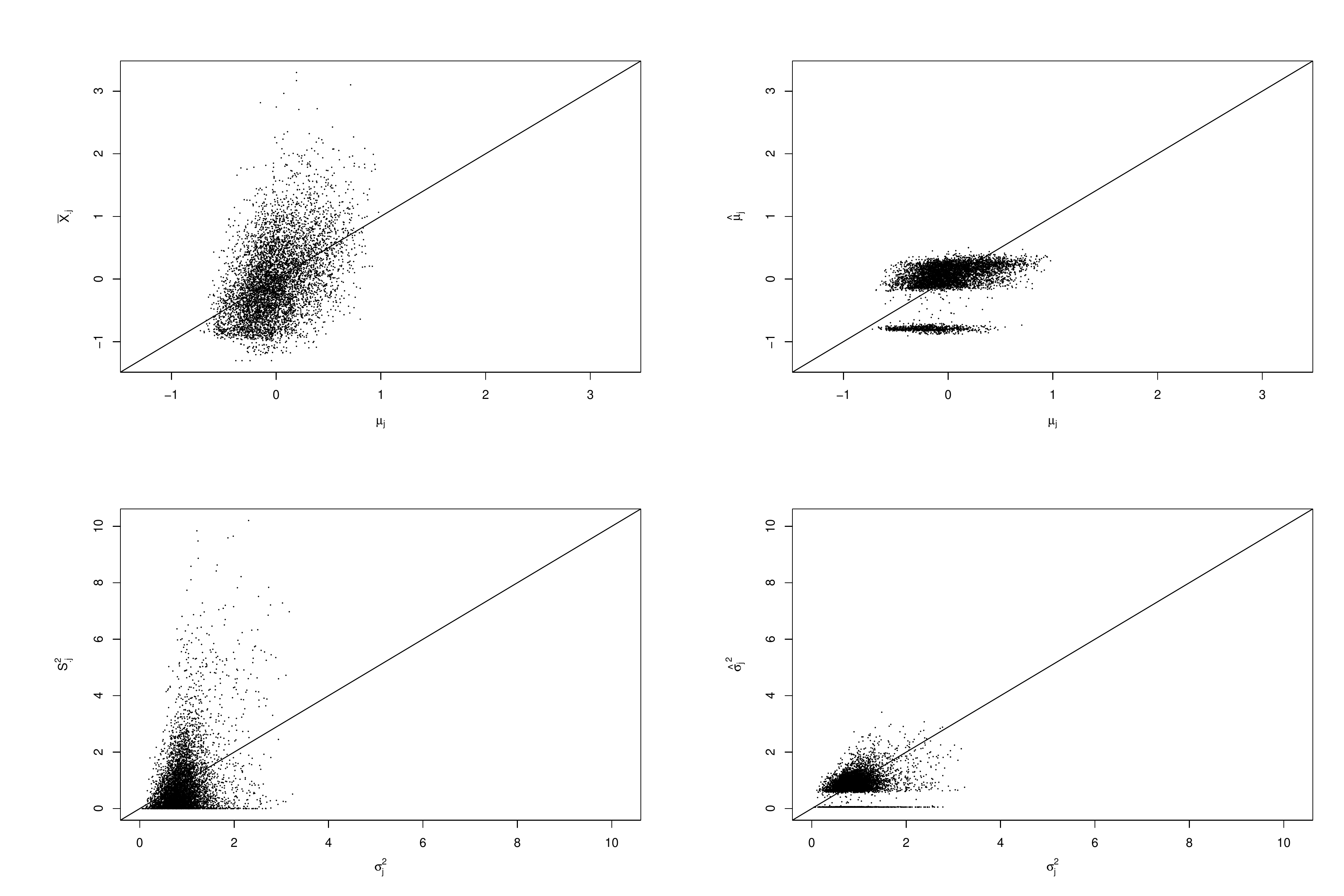}
\end{figure}

To better understand the nature of the data we provide the scatterplots in Figures \ref{fig:prostate1}-\ref{fig:prostate2}. We also compared our estimate with the sample means and variances from three columns.

\begin{figure}[hp]
\captionsetup{font=scriptsize}
\centering
\caption{Marginal kernel density estimates computed from $\mu_j$ and $\sigma_j^2$ based on all 50 columns of the data matrix.}\label{fig:prostate4} 
\includegraphics[width=12cm,height=5cm]{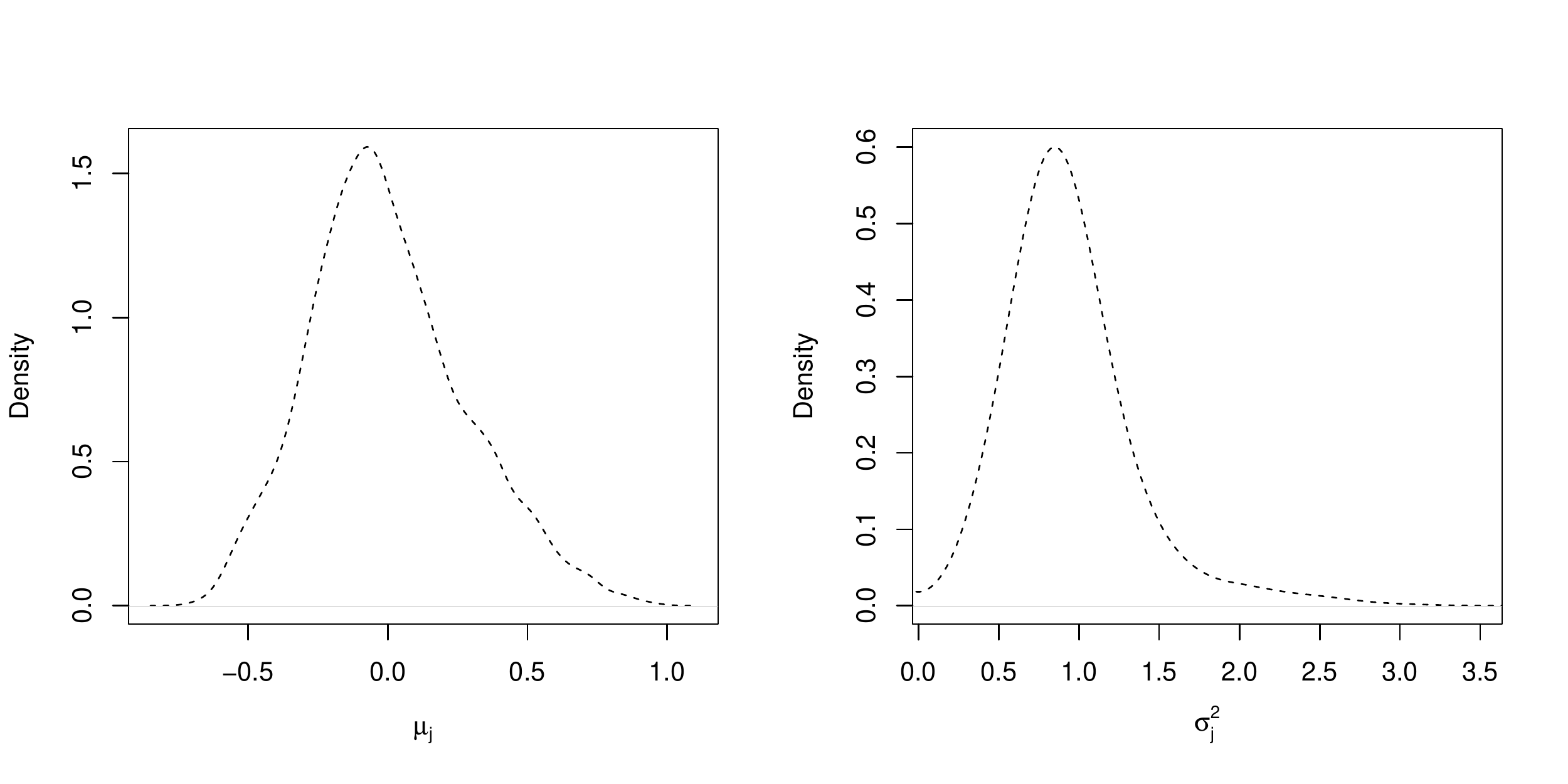}
\end{figure}

To compare different estimates we randomly chose 500 rows and 3 columns, computed estimates of means and variances using the various estimates, and replicated this process 100 times. Average squared error for each estimate was computed as in our simulation study. Table \ref{tab:T10} shows that, except for the SURE-based Double shrinkage estimators, all estimates were outperformed by \NGam. Figure \ref{fig:prostate4} shows that the densities of $\mu_j$ and $\sigma_j^2$ are well-approximated by normal and inverse gamma densities, respectively. When we force the mixture of normal-inverse gammas to select only one component, then this estimate performs comparably to SURE.M.Double for estimating both $\bm{\mu}$ and $\bm{D}$. For the other algorithms, replacing the unknown $\sigma_j^2$ with $S_{\cdot j}^2$ results in a loss in accuracy of those estimates. 

\begin{table}[H]
\scriptsize
\centering
\captionsetup{font=scriptsize}
\caption[T10]{Estimated average squared loss for $\bm{\mu}$ and $\bm{D}$ for different estimates from prostate-control data. Each table value is an average over $100$ replications. Each replication consists of 500 randomly chosen rows and $3$ randomly chosen columns from the original $6033 \times 50$ data matrix.}
\begin{tabular}{l cc }
\hline
\hline
Different & Different measures &\\
\cline{2-3}
estimates & Error in estimating $\mu_j$ & Error in estimating $\sigma_j^2$ \\ 
\hline
Sample Statistics 	&	0.2919	&	1.1695	\\
EBMLE.XKB 	&	0.1486	&	-	\\
EBMOM.XKB 	&	0.1446	&	-	\\
JS.XKB 	&	0.2787	&	-	\\
Oracle.XKB 	&	0.1108	&	-	\\
SURE.G.XKB 	&	0.1071	&	-	\\
SURE.M.XKB 	&	0.1175	&	-	\\
SURE.SG.XKB 	&	0.1445	&	-	\\
SURE.SM.XKB 	&	0.1682	&	-	\\
GL.WMBZ	&	0.1694	&	-	\\
GL.SURE.WMBZ	&	0.1802	&	-	\\
DGL.WMBZ	&	0.0690	&	-	\\
SURE.M.Double	&	0.0644	&	0.1458	\\
$N\Gamma^{-1}$ mixture 	&	0.1081	&	0.2284	\\
$N\Gamma^{-1}$ one component	&	0.0683	&	0.1653	\\
\hline
\hline\label{tab:T10}
\end{tabular}
\end{table}

\section{Summary} \label{sec:sec7}

Since Stein's work (\cite{stein1956inadmissibility}), there has been much progress in using shrinkage estimators of the mean of a high-dimensional normal vector. However, all of the previous work shrinks the sample means in the same direction. We have developed a very general algorithm which does not rely on the belief that all $\mu_j$ are of the same magnitude. Our estimate works by clustering sample means into different groups, and then shrinking an individual mean towards its corresponding group mean. Our algorithm outperforms SURE estimates when $\mu_j$ and $\sigma_j^2$ are dependent, and outperforms group-linear algorithms when $\mu_j$ and $\sigma_j^2$ are independent. When $\mu_j$ has a multimodal distribution or when $\sigma_j^2$ is unknown, our estimate based on mixtures of normal-inverse gamma distributions performed better than all the other estimates with which it was compared. Also, our approach allows us to estimate the joint density of $(\mu_j,\sigma_j^2)$, a problem which seems not to have been previously addressed. All code for our methodology is available online at \url{https://github.com/shyamalendusinha/mean_estimation}. 
 
\clearpage
\biblist

\end{document}